\documentclass[10pt,journal,compsoc]{IEEEtran}

\ifCLASSOPTIONcompsoc
  \usepackage[nocompress]{cite}
\else
  \usepackage{cite}
\fi
\usepackage{multirow}
\usepackage{array}
\usepackage{colortbl} 
\usepackage{float}
\usepackage{subcaption}
\usepackage{lscape}
\usepackage{amsmath}
\usepackage[ruled,vlined]{algorithm2e}
\usepackage{booktabs}
\usepackage{graphicx}
\usepackage{adjustbox}
\usepackage{xspace}
\usepackage{amssymb}
\usepackage{amsmath}
\usepackage{comment}
\ifCLASSINFOpdf
\else
\fi
\begin{document}
%
\title{Intrusion Detection in Heterogeneous Networks with Domain-Adaptive Multi-Modal Learning}
%
%
%
%

\author{
\IEEEauthorblockN{Mabin Umman Varghese, Zahra Taghiyarrenani}
\IEEEauthorblockA{}
\IEEEcompsocitemizethanks{\IEEEcompsocthanksitem  M. Umman Varghese and Z. Taghiyarrenani are with the Center for Applied Intelligent Systems Research (CAISR), Halmstad University, Halmstad, Sweden
E-mail: mabumm22@student.hh.se, zahra.taghiyarrenani@hh.se. 
}
}

%
%

\markboth{}%
{Intrusion Detection in Heterogeneous Networks with Domain-Adaptive Multi-Modal Learning}
%



\IEEEtitleabstractindextext{%
\begin{abstract}
Network Intrusion Detection Systems (NIDS) play a crucial role in safeguarding network infrastructure against cyberattacks. As the prevalence and sophistication of these attacks increase, machine learning and deep neural network approaches have emerged as effective tools for enhancing NIDS capabilities in detecting malicious activities. However, the effectiveness of traditional deep neural models is often limited by the need for extensive labelled datasets and the challenges posed by data and feature heterogeneity across different network domains. To address these limitations, we developed a deep neural model that integrates multi-modal learning with domain adaptation techniques for classification. Our model processes data from diverse sources in a sequential cyclic manner, allowing it to learn from multiple datasets and adapt to varying feature spaces. Experimental results demonstrate that our proposed model significantly outperforms baseline neural models in classifying network intrusions, particularly under conditions of varying sample availability and probability distributions. The model's performance highlights its ability to generalize across heterogeneous datasets, making it an efficient solution for real-world network intrusion detection.
\end{abstract}

}

\maketitle

\IEEEdisplaynontitleabstractindextext

%
\IEEEpeerreviewmaketitle

\ifCLASSOPTIONcompsoc
\IEEEraisesectionheading{\section{Introduction}\label{sec:introduction}}
\else
\section{Introduction}
\label{sec:introduction}
\fi
Network Intrusion Detection Systems (NIDS) are essential components of modern cybersecurity infrastructure, designed to monitor and analyze network traffic for signs of unauthorized access or malicious activity. These systems can be implemented as software or hardware solutions, with the primary goal of identifying and mitigating security breaches in real-time or near real-time. However, as cyber threats become more sophisticated, traditional signature-based detection methods struggle to keep pace, often failing to identify novel attacks that exploit previously unseen vulnerabilities. This has driven a shift towards machine learning (ML) and deep learning (DL) techniques, which can autonomously learn complex patterns from large volumes of network traffic, significantly enhancing the detection capabilities of NIDS \cite{ahmad2021network}, \cite{bib8}, \cite{sarhan2022towards}, \cite{dong2025cybersecurity}.

Despite their promise, deep neural network-based NIDS face significant challenges in real-world deployment. One of the primary hurdles is the phenomenon of domain shift, which arises from variations in network environments, including differences in protocols, traffic patterns, and feature distributions across different datasets. Traditional neural models typically assume a fixed input feature space, which makes them less effective when exposed to diverse real-world traffic, where feature types and dimensionalities can vary widely.

To address this, domain adaptation techniques are often employed to reduce the impact of domain shift by aligning feature distributions between different domains \cite{8566601},  \cite{taghiyarrenani2022domain},  \cite{fares2025deep},  \cite{shahid2025transfer},  \cite{ha2025tl},  \cite{taghiyarrenani2023multi}. Simultaneously, multi-modal learning aims to integrate information from diverse sources, each with different feature sets, enhancing model robustness \cite{farrukh2025xg}, \cite{ taghiyarrenani2022analysis}. However, in the context of NIDS, there is often significant feature overlap across different domains, such as common network statistics and shared traffic characteristics. This overlap presents a unique opportunity to combine both domain adaptation and multi-modal learning, leveraging shared features to enhance generalization while still capturing the unique characteristics of each domain. By integrating these approaches, a NIDS can learn more comprehensive, transferable representations to detect intrusions across varied network environments.

\begin{figure}[t]
    \centering
    \includegraphics[width=0.5\textwidth]{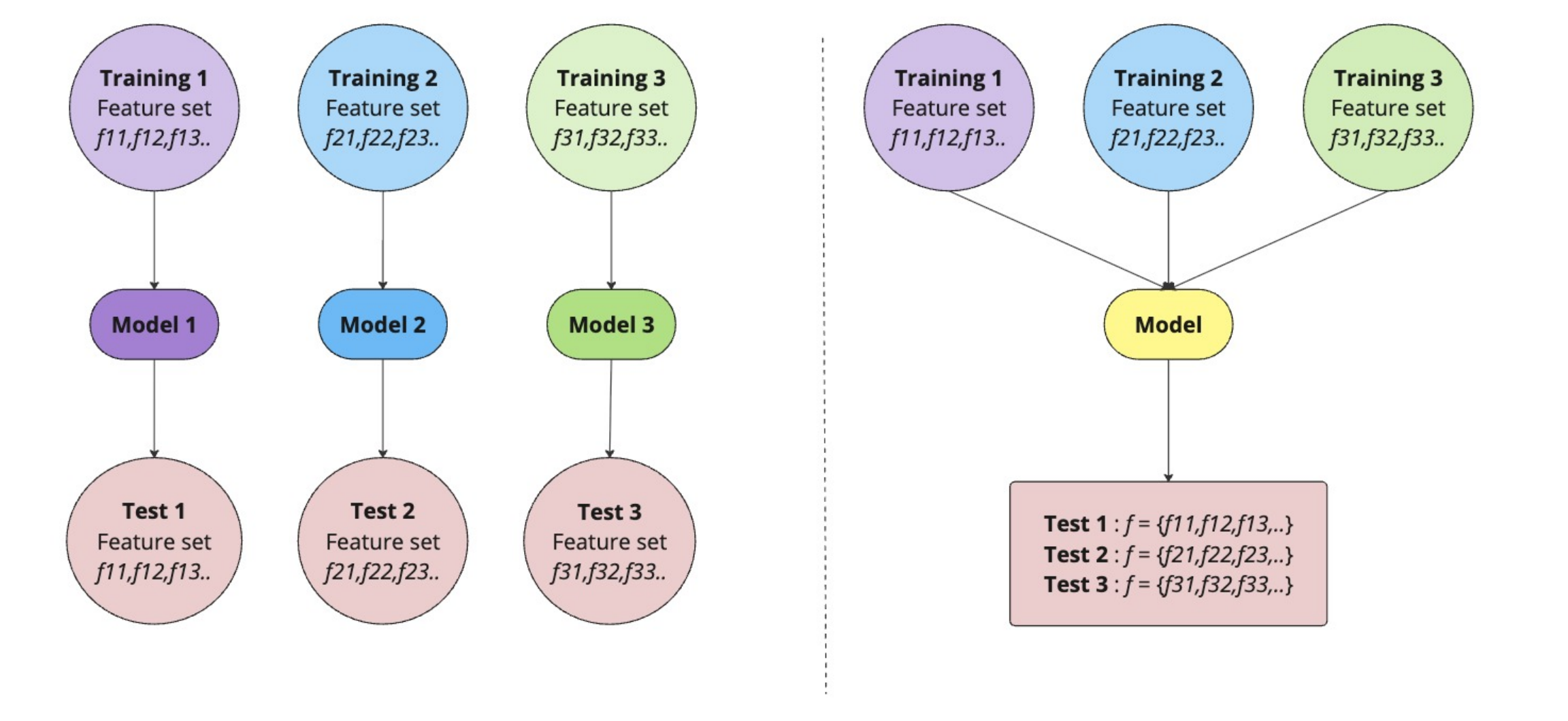}
    \caption{\textbf{Comparison of Traditional and Proposed NIDS Model -} On the left, it illustrates the traditional approach where models are trained independently on different datasets, each containing distinct feature sets. The right side presents the proposed NIDS model, which addresses feature heterogeneity by integrating features from multiple source datasets into a single adaptable model. This model is trained on diverse feature sets from different domains, enabling it to generalize across various scenarios and improve the detection of network attacks compared to the traditional approach.}
    \label{fig:Proposal}
\end{figure}

To address these challenges, we propose a novel deep learning-based NIDS framework that integrates domain adaptation and multi-modal learning to bridge the gap between isolated network domains. As illustrated in Figure~\ref{fig:Proposal}, traditional approaches often train separate models on individual datasets, each with a fixed feature set, limiting their ability to generalize across domains. In contrast, our proposed model is designed to learn from multiple, heterogeneous feature sets, enhancing its capacity to detect a broader spectrum of network intrusions.

The model is trained on multiple source datasets, each containing diverse feature sets, and validated on heterogeneous test data to assess its adaptability and generalization.

The primary contributions of this work include:\\
\textbf{Generalized Feature Adaptation:} Developing a deep neural architecture capable of integrating diverse feature sets from multiple source domains, enabling it to handle the feature variability often encountered in real-world network traffic.\\
\textbf{Improved Transferability:} Demonstrating the model's ability to generalize across heterogeneous datasets, improving detection accuracy even under varying network conditions.\\
\textbf{Comprehensive Evaluation:} Assessing the effectiveness of the proposed approach across multiple datasets, providing insights into its practical applicability for real-world NIDS deployments.



\section{Related Works}
Network security remains a critical concern for modern enterprises, as breaches can compromise organizational integrity, privacy, and trust. To mitigate these risks, many organizations deploy Network Intrusion Detection Systems (NIDS), leveraging a combination of hardware, software, and increasingly, cloud-based solutions. These systems are designed to identify and respond to potential threats by analyzing network traffic, detecting malicious activities, and alerting administrators to possible breaches. NIDS typically operate using three core approaches: signature-based detection, anomaly-based detection, and stateful protocol analysis.

\textbf{Signature-based detection} inspects network packets for predefined patterns or signatures that match known malicious or unauthorized activities. These signatures are crafted based on observed attack behaviors, including specific byte sequences, known exploit patterns, or protocol anomalies, allowing the system to identify and flag known threats with high precision \cite{wang2023deep},\cite{sarhan2022towards}.

\textbf{Anomaly-based detection} identifies potential intrusions by flagging deviations from established baseline behavior \cite{sarhan2022towards}.

\textbf{Stateful Protocol Analysis} monitors network traffic by comparing observed events against predefined profiles of legitimate protocol behavior. It tracks the state of communication sessions and identifies deviations that may indicate protocol misuse or unauthorized access \cite{vinay}.

The rapid growth of digital networks has significantly increased the volume and complexity of data, creating new challenges for cybersecurity. Traditional network security tools, such as firewalls, antivirus software, and signature-based intrusion detection systems (IDS), remain foundational but have critical limitations. Signature-based IDS, for instance, rely on predefined patterns to detect known threats, making them precise for familiar attacks but ineffective against novel, signature-evasive threats \cite{11}. This approach also requires frequent database updates, introducing latency and manual overhead.

To overcome these limitations, machine learning (ML) and deep learning (DL) have emerged as powerful tools for network intrusion detection. Unlike static, rule-based systems, ML and DL models can autonomously learn complex features from high-dimensional network traffic, capturing subtle patterns and correlations that might be missed by conventional methods \cite{10}, \cite{9}. This makes them particularly effective at detecting zero-day attacks and evolving cyber threats, which often evade traditional defenses.

Additionally, these AI-driven approaches offer several distinct advantages:
\\
\textit{Capturing Complex Attack Behaviors}: ML algorithms can identify subtle, often overlooked indicators of malicious activity by learning from diverse network traffic, significantly improving detection accuracy for sophisticated attacks \cite{2}, \cite{10}.
\\
\textit{Reduced Reliance on Signature Updates}: Unlike signature-based systems that require frequent updates to recognize new threats, ML models continuously adapt as they process new data, reducing the need for manual intervention and improving response times \cite{8386762}, \cite{11}.

As networks continue to expand, the demand for scalable, adaptive security solutions has become critical. According to the IBM 2023 Cyber Threat Report, the global average cost of a data breach has surged to USD 4.45 million, reflecting a 15\% increase over the past three years \cite{ibm-databreach}. This trend highlights the need for more advanced detection systems that can keep pace with the evolving threat landscape.


However, the success of ML-based Network Intrusion Detection Systems (NIDS) heavily depends on the availability of high-quality datasets for training and evaluation \cite{2}. Comparing the effectiveness of these models across different datasets is challenging, as each dataset often contains distinct and proprietary feature sets, reflecting the specific network environments or applications from which they were captured \cite{sarhan2022towards}. Network traffic features can vary significantly depending on the industry, domain, and devices involved, leading to differences in data distribution and feature representation. For instance, as shown in Figure 2, some datasets may share a few common features, but their distributions can vary widely, impacting model performance when deployed in different contexts. This variation emphasizes the need for standardized feature sets to enable fair and reliable evaluation of ML models across diverse network environments \cite{sarhan2022towards}. Additionally, obtaining real-world, labeled network flow datasets is challenging due to privacy concerns and proprietary restrictions, which often necessitates the use of synthetic datasets for training and testing \cite{3}. Figure \ref{fig:Proposal} illustrates this issue, highlighting both shared and unique feature sets across various network traffic datasets. This diversity in feature distributions and class ratios underscores the importance of domain adaptation techniques to ensure effective network intrusion detection across different environments.

\noindent Machine learning and deep learning have been increasingly used in NIDS because of their ability to process vast amounts of data and detect complex patterns that traditional methods may be limited \cite{2}, \cite{9}. However, the effectiveness of these models heavily depends on the quality and representativeness of the training data, which often varies significantly across different domains and datasets. Several studies have explored the use of ML techniques for network intrusion and traffic classification \cite{8386762},\cite{ahmad2021network}, revealing the complexities and challenges associated with training models under varying conditions. Most statistical models require re-modeling when data distribution changes, as network flow data obtained at one time may not match the distribution at another, regardless of the industry \cite{singla2020preparing}. This challenge highlights the need for models that can adapt to these changes without requiring constant retraining.\\

\noindent The discrepancy between datasets from different domains presents a significant challenge in training NIDS. In real industrial scenarios, network attack samples are often a minority compared to normal samples \cite{singla2020preparing} \cite{5}, and their characteristics may vary depending on the domain of origin \cite{6}. Traditional machine learning algorithms struggle with associating different feature distributions, especially when the feature sets and probability distributions across datasets are not common \cite{7}. \\

\noindent Advanced techniques like Transfer learning, domain adaptation, and multimodal learning can also be employed to address the challenges posed by handling heterogeneous datasets in NIDS. Transfer learning is particularly useful in scenarios where frequent updating of outdated data and recalibration are impractical. By leveraging pre-trained models, transfer learning can predict labels for future data samples, thereby reducing the need for extensive labeling efforts \cite{8566601}. Domain adaptation approaches like Domain Adversarial Neural Networks (DANN) facilitate domain invariance between known and variant attack data, improving the classification of modified attack patterns \cite{DANN9847041}, \cite{ding2019learning}. Manifold adaptation is a statistical method to map datasets to a common attribute space \cite{12}. Another method to address this challenge is the Maximum Mean Discrepancy (MMD), which has proven effective in aligning feature distributions between source and target domains \cite{wang2023deep} \cite{8566601} \cite{8}. This alignment enhances the system's ability to detect intrusions, ultimately adapting and improving generalization across diverse network environments. Additionally, feature-based transfer learning approaches, such as Clustering Enhanced Hierarchical Transfer Learning (CeHTL), enhance system robustness by improving the detection of new network attacks \cite{zhao2019transfer}. CeHTL determines the relationship between new and known attacks, automates the relevance finding between source and target domains, and assesses cluster similarities for better performance.\\

\noindent Deep learning-based algorithms, though effective, require large amounts of labeled data \cite{ahmad2021network}, which can be costly and time-consuming. Transfer learning and domain adaptation help mitigate these challenges by transferring knowledge from existing datasets, making them applicable even when feature sets differ across datasets. Furthermore, the application of Generative Adversarial Networks (GANs) in NIDS has shown promise in generating data for adversarial training, optimizing the detection models' performance even with varying feature sets \cite{singla2020preparing}.\\

\noindent Multimodal learning, which involves training models on diverse modalities, is another effective approach for enhancing NIDS performance. This learning method improves model resilience and flexibility by leveraging cross-modality learning and shared representation learning \cite{ngiam2011multimodal}. Our proposed approach utilizes a cyclic shared representation learning framework, which aims to determine whether the learned feature representation can capture correlations across multiple modalities, potentially leading to modality-invariant representations.

\section{Methodology} 


We propose an architecture that integrates multi-modal learning with a shared representation framework and domain adaptation to effectively handle heterogeneous data from different networks. These data may differ in their feature sets and are expected to have distinct probability distributions. The model processes input from different domains using private networks, each dedicated to learning domain-specific features. These private networks project the input features into a shared latent space, where we minimize the distributional discrepancy between domains.

\begin{figure*}[ht]
    \centering
    \includegraphics[width=0.7\linewidth]{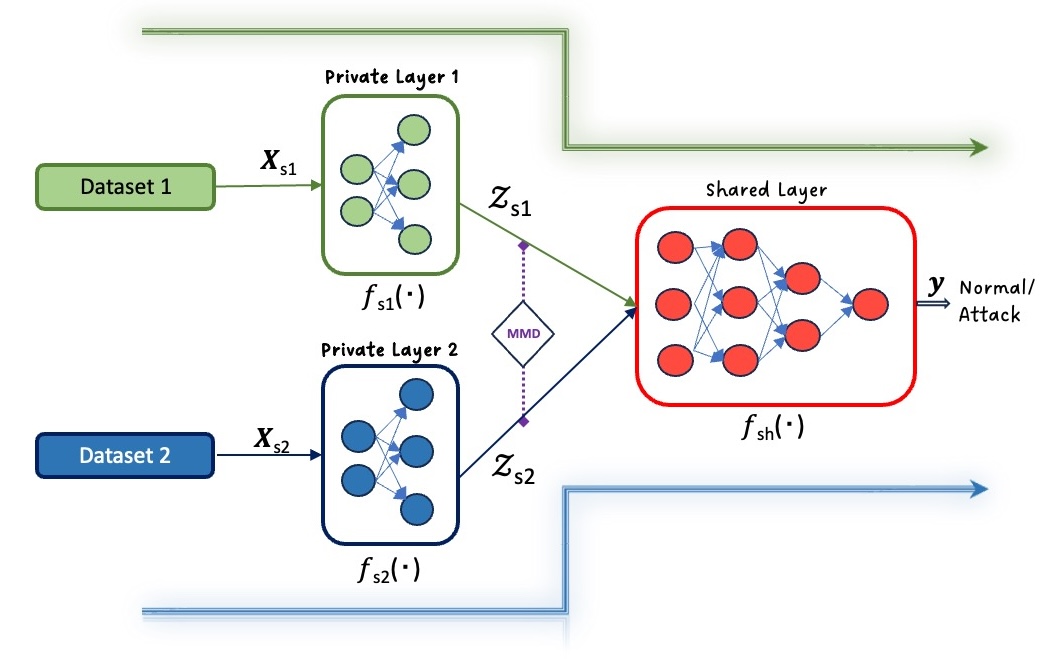}
    \caption{Model Architecture}
    \label{fig:Model_Arch}
\end{figure*}

The transformed features from the private networks are then forwarded to a shared network, which captures domain-invariant representations and performs the final classification. The model is trained using a combined loss function that includes class prediction and domain alignment. Training is conducted in a sequential cyclic manner, where each domain is used in turn to update the corresponding private network along with the shared network. This iterative process enables the model to gradually learn both shared and domain-specific features, improving generalization across diverse datasets. Although the approach is scalable to multiple datasets, we focus our discussion on a two-dataset scenario for clarity.
 Figure \ref{fig:Model_Arch} illustrates the proposed model architecture.

\noindent\textbf{Formal Definitions.}\\

\noindent The datasets \(D_{s1}\) and \(D_{s2}\) represent two heterogeneous datasets collected from different domains (networks), denoted as \(s_1\) and \(s_2\), respectively:
\[
D_{s1} = \left\{\left(x_i^{s_1}, y_i^{s_1}\right)\right\}_{i=1}^{n_{s1}}, \quad x_i^{s_1} \in \mathbb{R}^{d_{s1}}
\]
\[
D_{s2} = \left\{\left(x_j^{s_2}, y_j^{s_2}\right)\right\}_{j=1}^{n_{s2}}, \quad x_j^{s_2} \in \mathbb{R}^{d_{s2}}
\]
Here, \(x\) denotes the input features and \(y\) denotes the corresponding class labels. We assume that the datasets have different feature dimensions, i.e., \(d_{s1} \neq d_{s2}\).

The proposed model consists of two private networks, \(f_{s1}(\cdot)\) and \(f_{s2}(\cdot)\), along with a shared network \(f_{sh}(\cdot)\). The private networks are responsible for learning feature representations specific to each dataset, effectively capturing domain-specific characteristics and projecting them into a common latent space \(Z\):

\begin{equation}
Z_{s1} = f_{s1}(X_{s1})
\end{equation}

\begin{equation}
Z_{s2} = f_{s2}(X_{s2})
\end{equation}

The latent representations obtained from the private networks are then passed to the shared network \(f_{sh}(\cdot)\), which produces the final output prediction \(y\).

A key concept of the proposed model is the adaptation of the transformed features before they are passed to the shared network \(f_{sh}(\cdot)\). This adaptation addresses domain shift by computing the Maximum Mean Discrepancy (MMD) between the transformed feature sets \(Z_{s1}\) and \(Z_{s2}\), obtained from the respective private networks \(f_{s1}(\cdot)\) and \(f_{s2}(\cdot)\). By projecting the features onto a shared latent space, the model learns domain-invariant representations, aligning the distributions of inputs from different domains to capture meaningful and generalizable patterns. This enables the model to focus on shared representations and minimize domain-specific discrepancies, thereby improving its ability to transfer knowledge across domains.

We denote the transformed feature sets as \(Z_{s1} \in \mathbb{R}^{N \times D}\) and \(Z_{s2} \in \mathbb{R}^{M \times D}\), where \(N\) and \(M\) are the number of samples from each domain, and \(D\) is the dimensionality of the latent space.

The MMD computation is defined as:

\begin{equation}
\operatorname{MMD}\left(Z_{s1}, Z_{s2}\right) = \left\| \frac{1}{N} \sum_{i=1}^N \phi\left(z_i^{s_1}\right) - \frac{1}{M} \sum_{j=1}^M \phi\left(z_j^{s_2}\right) \right\|^2
\end{equation}

\noindent where \(z_i^{s_1} \in Z_{s1}\) and \(z_j^{s_2} \in Z_{s2}\) are the transformed outputs of the respective private networks. To capture non-linear relationships between distributions, we employ a Gaussian kernel function. The mapping \(\phi(z): Z \rightarrow \mathcal{H}\) embeds the feature into a universal Reproducing Kernel Hilbert Space (RKHS) \(\mathcal{H}\) \cite{8}, enabling a flexible and powerful similarity measure between samples from different domains.

The empirical MMD loss between the two heterogeneous datasets is computed as:

\begin{equation}
\label{eq:mmd}
\begin{aligned}
\mathrm{MMD}
&= \frac{1}{N^2}
  \sum_{i=1}^{N} \sum_{j=1}^{N}
    k\bigl(z_i^{s_1}, z_j^{s_1}\bigr) \\
&\quad + \frac{1}{M^2}
  \sum_{i=1}^{M} \sum_{j=1}^{M}
    k\bigl(z_i^{s_2}, z_j^{s_2}\bigr) \\
&\quad - \frac{2}{NM}
  \sum_{i=1}^{N} \sum_{j=1}^{M}
    k\bigl(z_i^{s_1}, z_j^{s_2}\bigr)
\end{aligned}
\end{equation}

\noindent where \(k(z_i^{s_1}, z_j^{s_1})\) denotes the kernel function that computes the inner product \(\langle \phi(z_i^{s_1}), \phi(z_j^{s_1}) \rangle\) in the RKHS. Accordingly, the first loss function used for training the model is defined as:

\begin{equation}
L_{\mathrm{MMD}} = \operatorname{MMD}\left(Z_{s1}, Z_{s2}\right)
\end{equation}

The transformed feature set \(Z_{s1}\) is then passed to the shared network \(f_{sh}(\cdot)\), which acts as a classifier to predict the output. Accordingly, we employ a second loss function: the cross-entropy loss, defined as:

\begin{equation}
L_{\mathrm{CE}} = -\sum_{i=1}^{\text{size}} y_i \cdot \log \hat{y}_i, \quad \text{where size} \in \{N, M\}
\end{equation}

\noindent In the first phase of training, we optimize a combined loss function that integrates the cross-entropy loss \(L_{\mathrm{CE}}\) and the Maximum Mean Discrepancy (MMD) loss \(L_{\mathrm{MMD}}\) to update the parameters of both the shared network \(f_{sh}(\cdot)\) and the private network \(f_{s1}(\cdot)\). The combined loss is expressed as:

\begin{equation}
\label{eq:conmbined_loss}
\text{Loss} = \alpha \cdot L_{\mathrm{CE}} + \beta \cdot L_{\mathrm{MMD}}
\end{equation}

\noindent where \(\alpha\) and \(\beta\) are weighting coefficients that balance the contribution of the two losses.

\section{Experiments}
\label{ch:Dataset_study} 

\subsection{Experimental Setups}

\noindent We construct 20 cross-domain datasets using the datasets listed in Table \ref{tab:Datasets}, leveraging four publicly available datasets \cite{zoppi2022towards}. These datasets cover a range of attack categories, including HTTP Denial of Service, Web Application Attacks, and Brute Force Attacks, and provide information on the percentage distribution of attack samples in each dataset. All datasets share a common feature set of 77 attributes, with samples labeled as either \textit{NORMAL} or by their respective attack type. The feature sets are detailed in Table \ref{Dataset_Feature}.

Features and labels classify the samples as either \textit{NORMAL} or \textit{ATTACK}, encompassing various attack categories (e.g., HTTP denial of service, web application attacks, brute force attacks, etc.). Each of the four datasets contains distinct types of attack samples (see Table \ref{tab:Datasets}). For classification purposes, we consolidate all attack types into a single class labeled "ATTACKS," while retaining "NORMAL" for the remaining samples.

\begin{table}[ht]
\centering
\caption{Datasets used for the proposed exploration\cite{zoppi2022towards}}
\resizebox{\columnwidth}{!}{%
\begin{tabular}{@{}llllll@{}}
\toprule
Sr\# & Dataset & Datapoints & Features & \begin{tabular}[c]{@{}l@{}}Types of\\  Attacks\end{tabular} & Attacks(\%) \\ \midrule \midrule
1 & CIC-IDS2017 & 152,055 & 77 & 5 & 79.7 \\
2 & CIC-IDS2018 & 199,997 & 77 & 8 & 26.2 \\
3 & SDN20       & 205,167 & 77 & 5 & 66.6 \\
4 & ANDMAL2017  & 100,521 & 77 & 4 & 15.5 \\ \bottomrule
\end{tabular}%
}
\label{tab:Datasets}
\end{table}

\subsection{Cross-domain datasets and Evaluation setups}\label{sec:options}

To evaluate our methodology, we introduce 20 different cross-domain datasets with unique feature sets, which were randomly selected from each dataset. Table \ref{tab:Com_table} in the Appendix shows the description of the different cross-domain datasets used for the evaluation. 
For evaluation, we design the following experiment setups:\\
\textbf{Setup 1:} Using one single dataset, we construct different cross-domain datasets by splitting the feature set.\\
\textbf{Setup 2:} We construct cross-domain datasets using two different datasets. \\
\textbf{Setup 3:} We construct cross-domain datasets using three datasets.

In addition, to validate the robustness of the proposed model, it is essential to consider how it performs under varying sample variability and distribution. To assess these factors, we design the following experiments using the cross-domain datasets in \textbf{setup 1}.

\begin{itemize}
    \item We take different combinations of feature sets [5-5], [15-15], [5-20], [30-5] to see how the difference in model performance with change in feature dimensionality.
    \item We sub-sample the second subset of all the combination subsets to 10\%, 50\%, 75\%, 100\% of its original size (table \ref{tab:subset_sample_ratio}) while keeping the first subset unchanged. 
\end{itemize}  

 For all of the constructed cross-domain datasets, standard preprocessing techniques were applied using the same pipeline. Initially, all duplicates and null values were removed. The dataset was split into training, validation, and testing sets, with stratified sampling to preserve the label distribution. The training set was further balanced using two resampling techniques: SMOTE (Synthetic Minority Over-sampling Technique) to increase the number of minority class samples, followed by RandomUnderSampler() to reduce the number of majority class samples.  The features were then transformed using a preprocessing pipeline that employed StandardScaler() to scale the features.

\begin{table}[t!]
\centering
\caption{Overview of Datasets with sub-sampled sample sizes (10\%, 50\%, 75\%)}
\begin{tabular}{l|c|c|c|c }
\toprule
\multicolumn{1}{c|}{\textit{\textbf{Dataset Name}}} &
  \textit{\textbf{Total datapoints}} &
  \textit{\textbf{10\%}} &
  \textit{\textbf{50\%}} &
  \textit{\textbf{75\%}} \\ \midrule \midrule
CIC-IDS2017     & 152,055 & 15,206 & 76,028 & 114,041 \\ \midrule
CIC-IDS2018  & 199,997 & 20,000 & 99,998 & 149,998 \\ \midrule
SDN20           & 205,167 & 20,517 & 102,584 & 153,875 \\ \midrule
AndMal2017   & 100,521 & 10,052 & 50,260 & 75,391 \\ \bottomrule
\end{tabular}

\label{tab:subset_sample_ratio}
\end{table}

\subsection{Results}
We show the results of the experiments in different plots, as illustrated in Figure 4-23. The plots illustrate the comparison of the metric scores, Accuracy, F1-score, Precision, and Recall, between the baseline and proposed model on both the derived subsets. The plot adapts a color shading to highlight the comparison between the evaluation metrics, with lighter shades representing the baseline scores and the corresponding darker shades representing the proposed model's performance.\\
In the following section, we analyze each combination set from Table \ref{tab:Com_table} (Sr\# 1-20), considering the factors above, and evaluate and analyze the proposed model performance with a change in sample variability and feature dimensionality.\\

\subsection{Setup 1}
\subsubsection{Evaluation of combinatorial sets derived from CIC-IDS2017 dataset}

As mentioned above, we randomly extract subsets of features from the CIC-IDS2017 dataset, forward them to the proposed model, and evaluate detection rate scores against their baseline classifier model.\\

\begin{figure*}[ht]
    \centering
    \includegraphics[width=1\linewidth,height=0.4\linewidth,keepaspectratio=false]{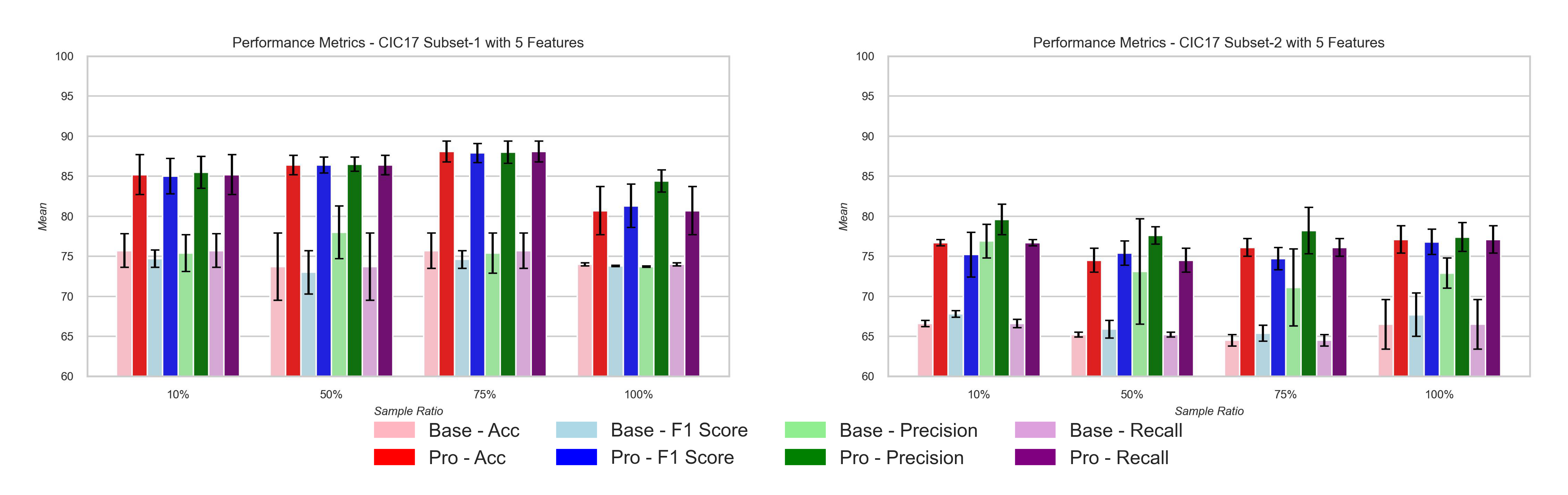}
    \caption{Evaluation of two domains with 5 unique feature sets from CIC-IDS2017 - Table \ref{tab:Com_table} (Sr\# 6)}
    \label{fig:cic17_1}
\end{figure*}

Figure \ref{fig:cic17_1}  evaluates the performance of the baseline and proposed model across different sample ratios (10\%, 50\%, 75\%, and 100\%) for two CIC-IDS2017 subsets (Subset-1 and Subset-2), each with 5 unique features. On observation, it is evident that the proposed model consistently outperforms the baseline in all metrics, with higher mean values, indicating a more reliable performance. Even with a smaller sample ratio of 10\%, the proposed model shows significant improvement over the baseline, which suggests that the distribution alignment effectively leverages the information from both subsets, leading to a better generalization even with limited data. As the sample size increases, the performance gap between the proposed model and the baseline remains, further highlighting the effective adaptation of multiple domains. This trend across all sample ratios showcases the proposed model's robustness and adaptability, demonstrating its effectiveness in handling variability and effectively learning from both datasets simultaneously.\\

\begin{figure*}[ht]
    \centering
    \includegraphics[width=1\linewidth,height=0.4\linewidth,keepaspectratio=false]{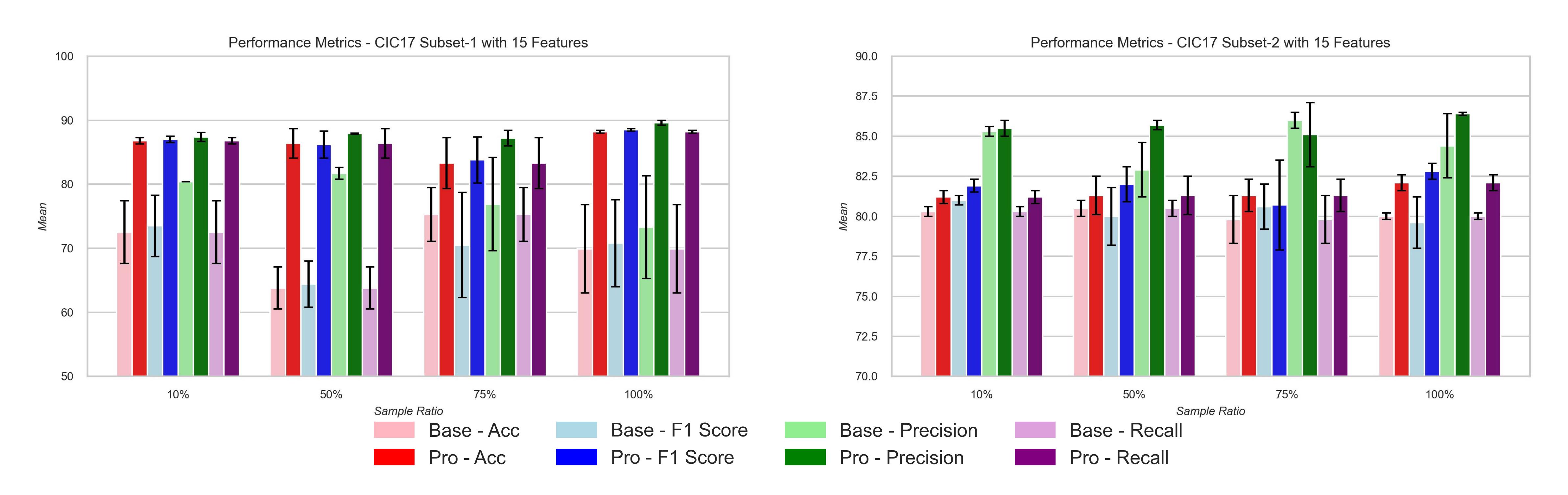}
    \caption{Evaluation of two domains with 15 unique feature sets from CIC-IDS2017  - Table \ref{tab:Com_table} (Sr\# 7)}
    \label{fig:CIC17_2}
\end{figure*}

Next, we study a new subset with 15-15 features from the CIC-IDS2017 dataset. Both subsets show consistent performance with minimal variation across different sample ratios. Compared to the earlier plot, where only 5 features were used, the overall mean performance metrics, including accuracy, F1 score, precision, and recall, show a further improvement in the detection of samples, suggesting that additional features contribute to a more informative representation of the data. The proposed model maintains higher mean values than the baseline model, regardless of the sample ratio. Notably, with the increased number of features, the performance metrics are more stable, with reduced variability, as shown in figure \ref{fig:CIC17_2}. This highlights the ability of the proposed model to effectively align and generalize the distributions between the two datasets, thereby achieving improved performance.\\

\begin{figure*}[ht]
    \centering
    \includegraphics[width=1\linewidth,height=0.4\linewidth,keepaspectratio=false]{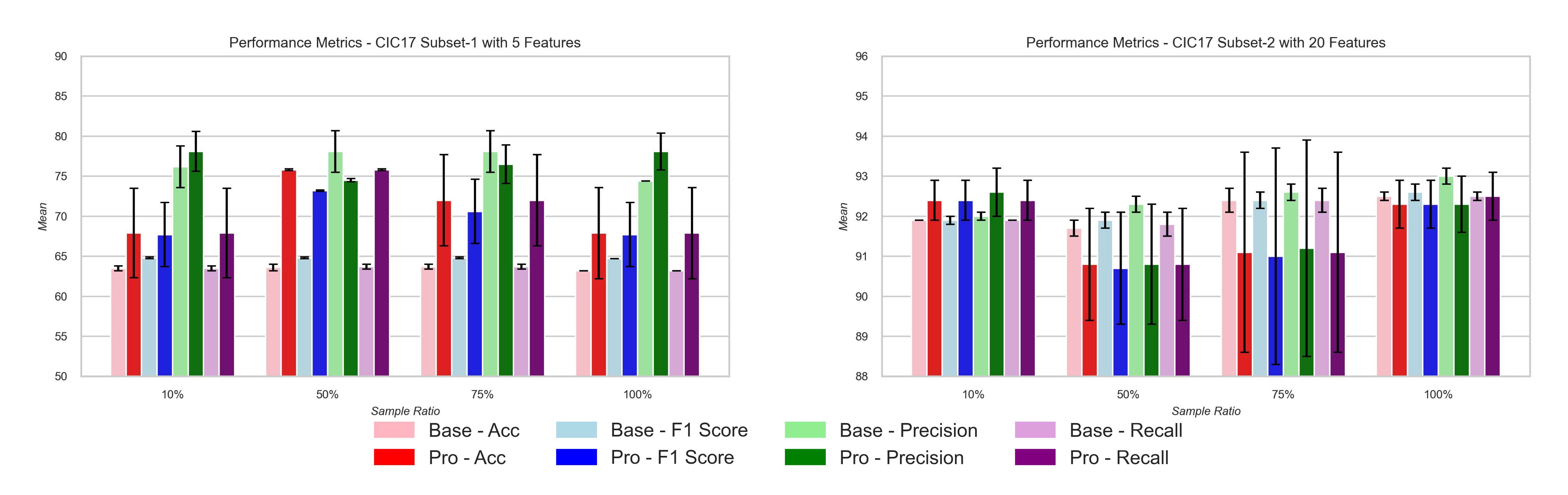}
    \caption{Evaluation of two domains with 5 and 20 unique feature sets from CIC-IDS2017 - Table \ref{tab:Com_table} (Sr\# 8)}
    \label{fig:cic17_3}
\end{figure*}

Following this, we conducted training using a new Subset-1 (5 features) and Subset-2 (20 features) (Figure \ref{fig:cic17_3}). On observing the subset 1 evaluation, the baseline scores show considerable variability, particularly when using only 10\% of the sample data, indicating the model's struggle with limited feature representation. On the other hand, the proposed model manages to improve the performance metrics consistently, indicating the model's ability to generalize well even with fewer features and a variable sample ratio. While with test subset 2 both the baseline and the proposed model achieve high and very similar performance metrics across all evaluation metrics. However, the proposed model still maintains a marginal improvement over the baseline, emphasizing its adaptability in aligning data distributions effectively. We can observe that the proposed model effectively learns from and adapts to the available data, benefiting from the shared learning process between Subset-1 and Subset-2.\\

\begin{figure*}[ht]
    \centering
    \includegraphics[width=1\linewidth,height=0.4\linewidth,keepaspectratio=false]{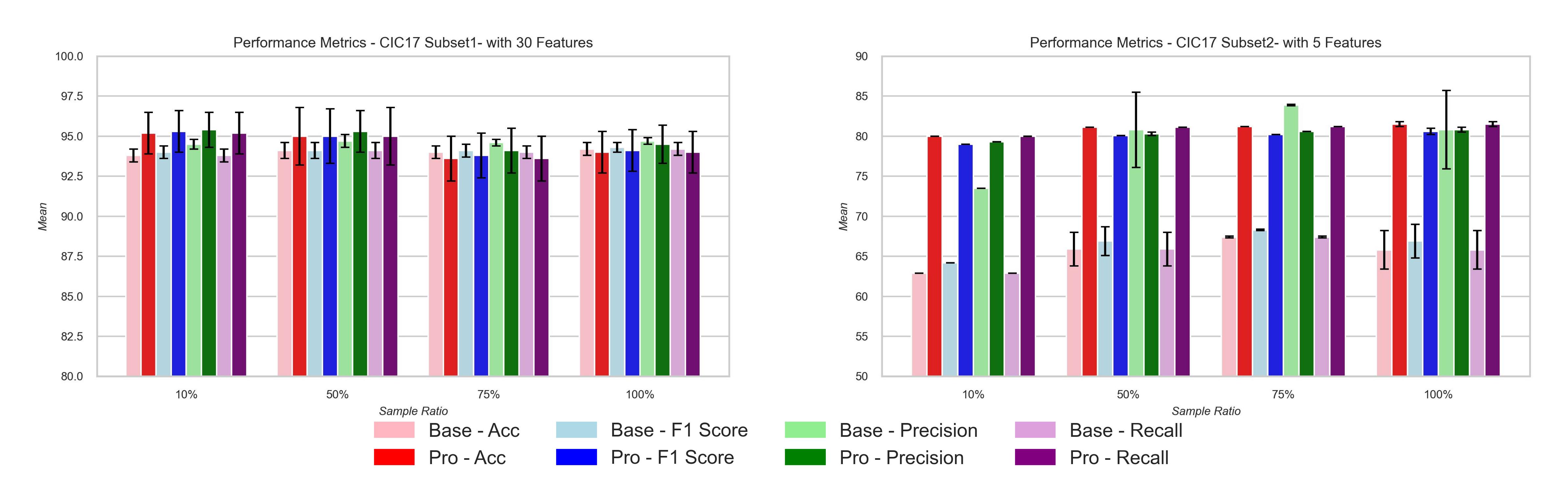}
    \caption{Evaluation of two domains with 30 and 5 unique feature sets - Table \ref{tab:Com_table} (Sr\# 9)}
    \label{fig:cic17_4}
\end{figure*}

Next, we conducted another training using a new set of subsets with 30-5 feature sets from the CIC-IDS2017 dataset. The results, as illustrated in figure \ref{fig:cic17_4}, indicate that the proposed model's ability to adapt and learn from multiple datasets is highly effective, particularly when there is an imbalance in feature dimensionality between the subsets. For Subset-1, with its 30 features, both the baseline and proposed model perform well, On the other hand, for Subset-2, which only has 5 features, the baseline model struggles due to the limited feature sets, whereas the proposed model's superior performance to learn from feature spaces from both the subsets effectively and compensate for the lower dimensionality. This adaptability, particularly in handling scenarios where one dataset has much higher feature sets than the other, highlights the strength of the proposed model in heterogeneous environments.

\subsubsection{Evaluation of combinatorial sets derived from CIC-IDS2018 dataset}

\begin{figure*}[ht]
    \centering
    \includegraphics[width=1\linewidth,height=0.4\linewidth,keepaspectratio=false]{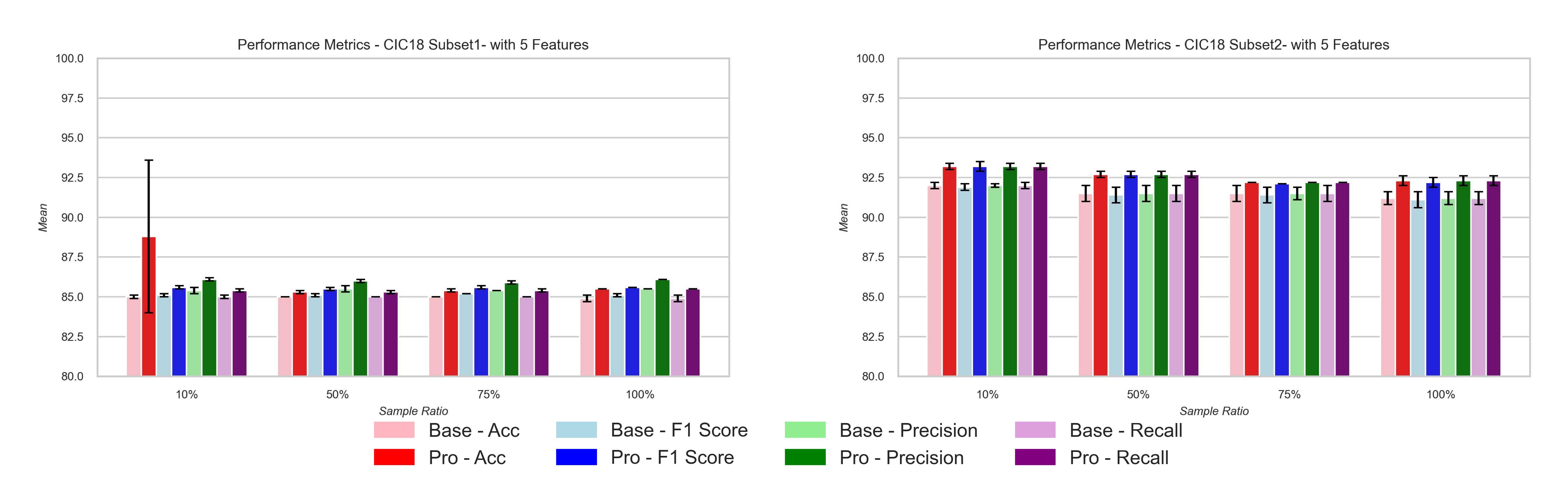}
    \caption{Evaluation of two domains from CIC18 dataset with 5 and 5 unique feature sets - Table \ref{tab:Com_table} (Sr\# 2)}
    \label{fig:cic18_1}
\end{figure*}

In the evaluation of a new dataset, CIC18, both Subset-1 and Subset-2 are assessed using 5 unique feature sets, as shown in figure \ref{fig:cic18_1}. The performance metrics reveal consistently high mean values across different sample ratios for both the baseline and proposed models. The proposed model generally achieves slightly higher performance than the baseline across all metrics. This demonstrates the effectiveness of the proposed model in maximizing classification performance even with a limited number of features. In Subset-1, the proposed model still maintains a slight advantage in terms of higher metric scores and reduced variability, while Subset-2 shows almost indistinguishable performance between the baseline and proposed models, especially for larger sample sizes (50\%, 75\%, and 100\%). Overall, these findings support the efficacy of the proposed model in heterogeneous scenarios with feature-limited datasets, achieving consistent performance metrics across varied sample sizes.

\begin{figure*}[ht]
    \centering
    \includegraphics[width=1\linewidth,height=0.4\linewidth,keepaspectratio=false]{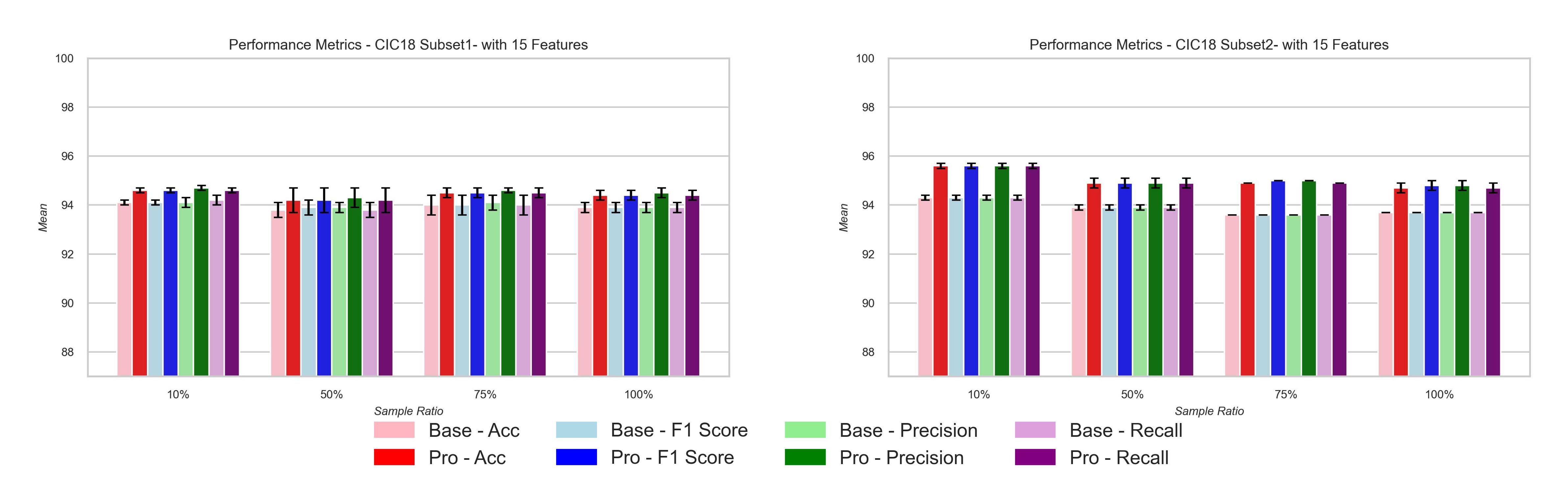}
    \caption{Evaluation of two domains from CIC18 dataset with 15 and 15 unique feature sets - Table \ref{tab:Com_table} (Sr\# 3)}
    \label{fig:cic18_2}
\end{figure*}

Next, we evaluate the CIC18 dataset with Subset-1 and Subset-2, both having 15 unique features as shown in figure \ref{fig:cic18_2}. The evaluation plots show that the proposed model consistently handles sample variability well across different sample ratios, maintaining high metric scores. The proposed model demonstrates slight improvements over the baseline model. As the number of features increases, the model adapts to distribution changes, and its performance remains stable with minimal variability, showing an upward trend compared to the baseline scores. This suggests that the proposed model adapts well to different sample sizes and also leverages the total number of features and distribution alignment to achieve better overall classification performance.

\begin{figure*}[ht]
    \centering
    \includegraphics[width=1\linewidth,height=0.4\linewidth,keepaspectratio=false]{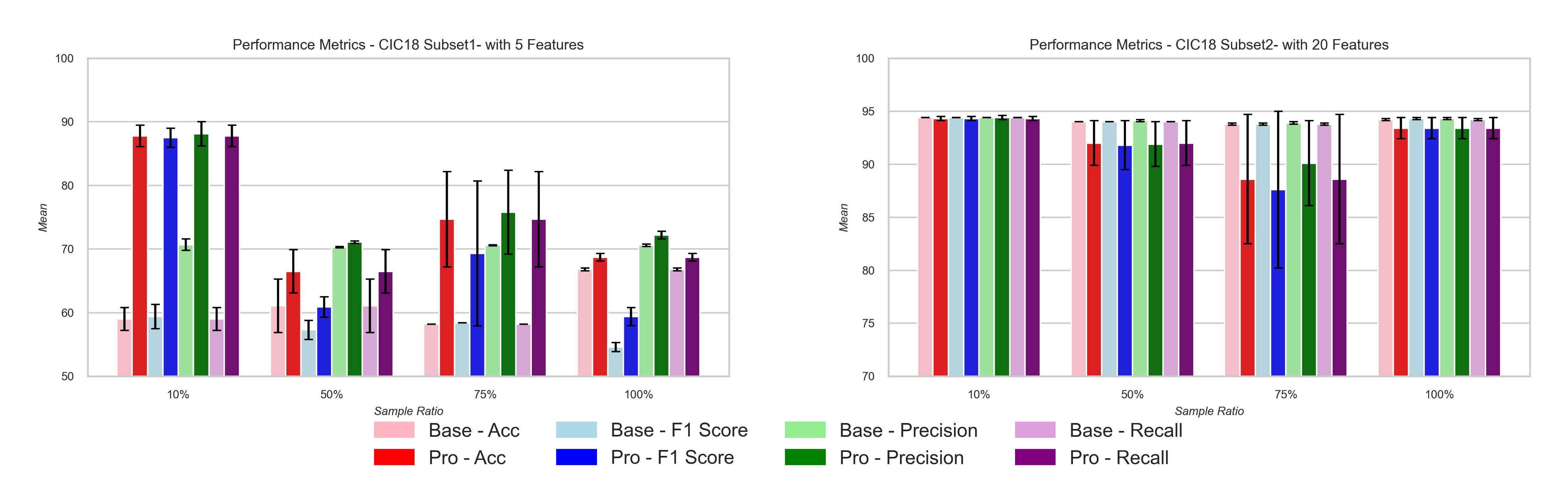}
    \caption{Evaluation of two domains from CIC18 dataset with 5 and 20 unique feature sets - Table \ref{tab:Com_table} (Sr\# 4)}
    \label{fig:cic18_3}
\end{figure*}

\begin{figure*}[ht]
    \centering
    \includegraphics[width=1\linewidth,height=0.4\linewidth,keepaspectratio=false]{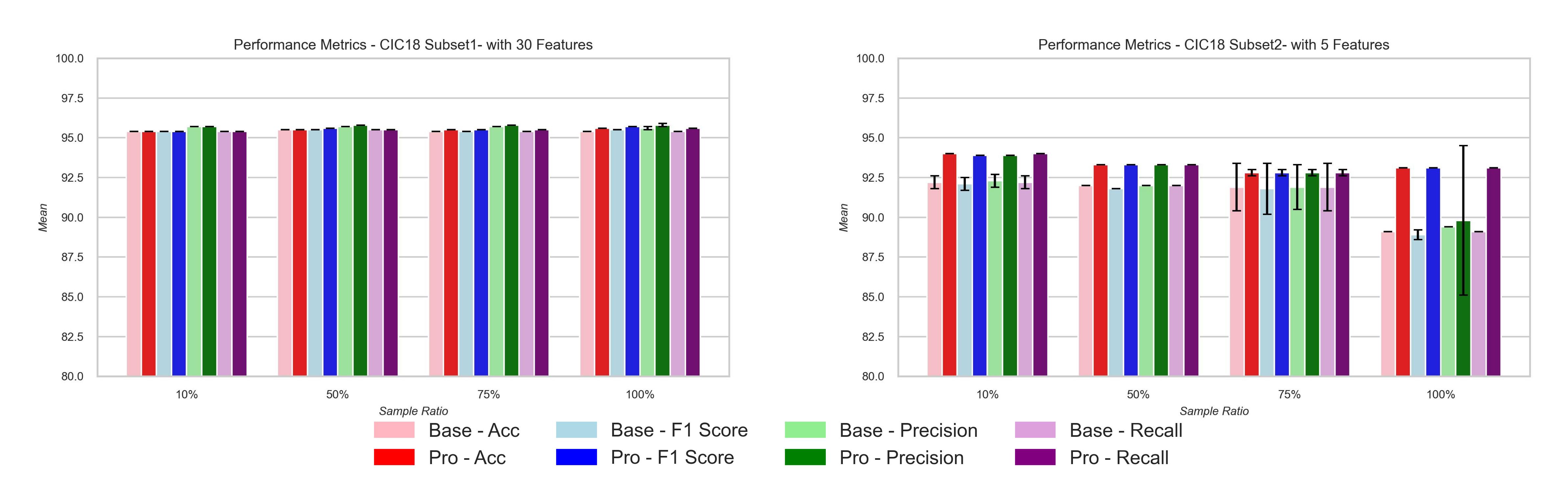}
    \caption{Evaluation of two domains from CIC18 dataset with 30 and 5 unique feature sets - Table \ref{tab:Com_table} (Sr\# 5)}
    \label{fig:cic18_4}
\end{figure*}

We then take into account two unique subsets with an imbalance in the total number of features. This can be seen in figure \ref{fig:cic18_3} and \ref{fig:cic18_4}; the proposed model demonstrates a clear advantage over individual baseline classifiers by effectively learning from both the subsets with a different number of features, one with a high number of features and one with a low number (Sr\#4, Sr\#5 in Table \ref{tab:Com_table} in Appendix). The proposed model is able to align the distributions of these heterogeneous subsets, leading to improved classification performance for both test datasets compared to their respective baseline models.

When one subset has more features, the proposed model benefits significantly by leveraging this richer information, which results in improved scores across all metrics, even for the subset with fewer features. The distribution alignment allows the proposed model to compensate for the disparity in feature availability, thereby enhancing learning from both subsets and providing better results, regardless of variable sample sizes. This demonstrates the robustness of the proposed model in handling feature imbalance and aligning distributions to improve overall classification.

\subsubsection{Evaluation of combinatorial sets derived from SDN20 dataset}

\begin{figure*}[ht]
    \centering
    \includegraphics[width=1\linewidth,height=0.4\linewidth,keepaspectratio=false]{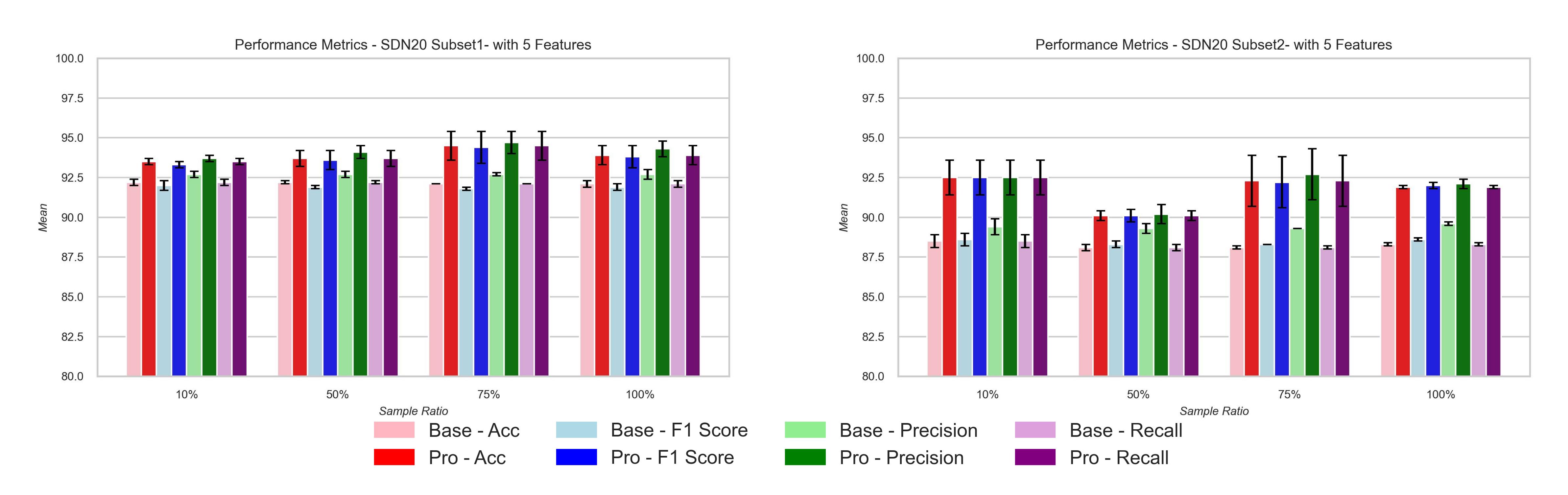}
    \caption{Evaluation of two domains from SDN20 dataset with 5 and 5 unique feature sets - Table \ref{tab:Com_table} (Sr\# 10)}
    \label{fig:sdn20_1}
\end{figure*}

In evaluating the SDN20 dataset, where both Subset-1 and Subset-2 have 5 unique features each, as shown in figure \ref{fig:sdn20_1}, the proposed model demonstrates consistent improvement over the baseline across all metrics at different sample ratios. The proposed model effectively aligns the distributions and learns from both subsets, resulting in enhanced performance on both test datasets despite the limited number of features. The quality of the selected features also plays a critical role in this improvement, suggesting a better classification of both normal and attack instances. Furthermore, the proposed model shows the capability to handle feature variability, as indicated by the reduced variability observed in the smaller error bars.\\

\begin{figure*}[ht]
    \centering
    \includegraphics[width=1\linewidth,height=0.4\linewidth,keepaspectratio=false]{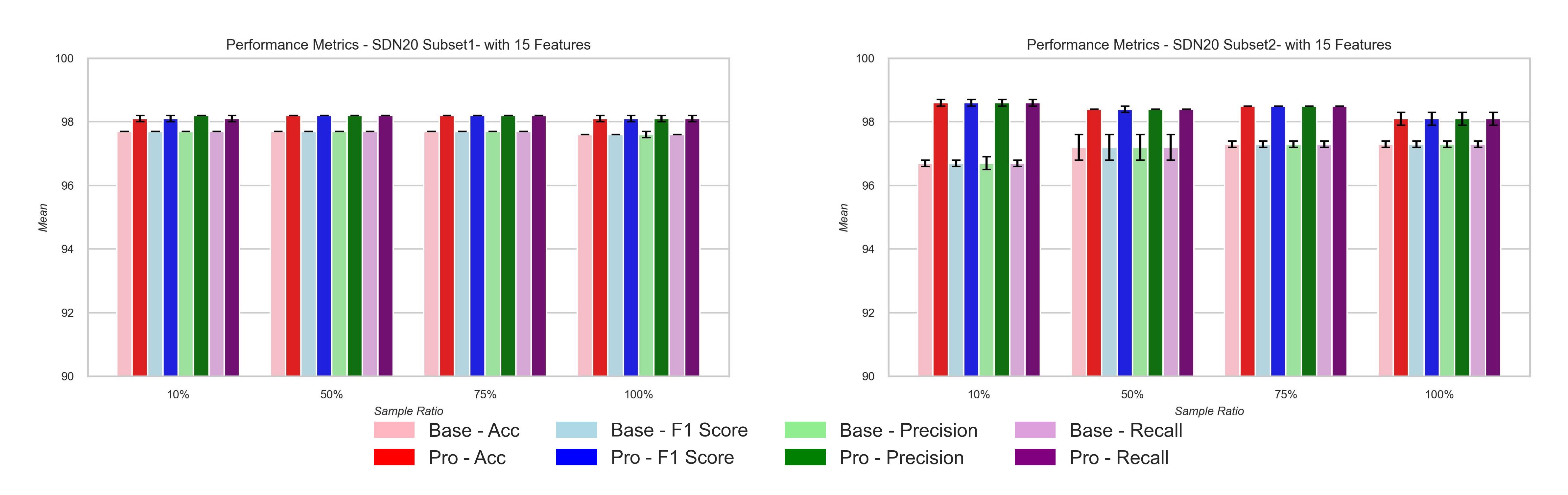}
    \caption{Evaluation of two domains from SDN20 dataset with 15 and 15 unique feature sets - Table \ref{tab:Com_table} (Sr\# 11)}
    \label{fig:sdn20_2}
\end{figure*}

Similarly, with the CIC-IDS2018 dataset, we now evaluate the SDN20 dataset, with both Subset-1 and Subset-2 having 15 unique features each (Figure \ref{fig:sdn20_2}). Here, the proposed model also demonstrates slight improvement over the baseline across all metrics at different sample ratios. The proposed model effectively aligns the distributions from these increased sets of feature sets and learns from both subsets, leading to enhanced performance on both test datasets.\\

\begin{figure*}[ht]
    \centering
    \includegraphics[width=1\linewidth,height=0.4\linewidth,keepaspectratio=false]{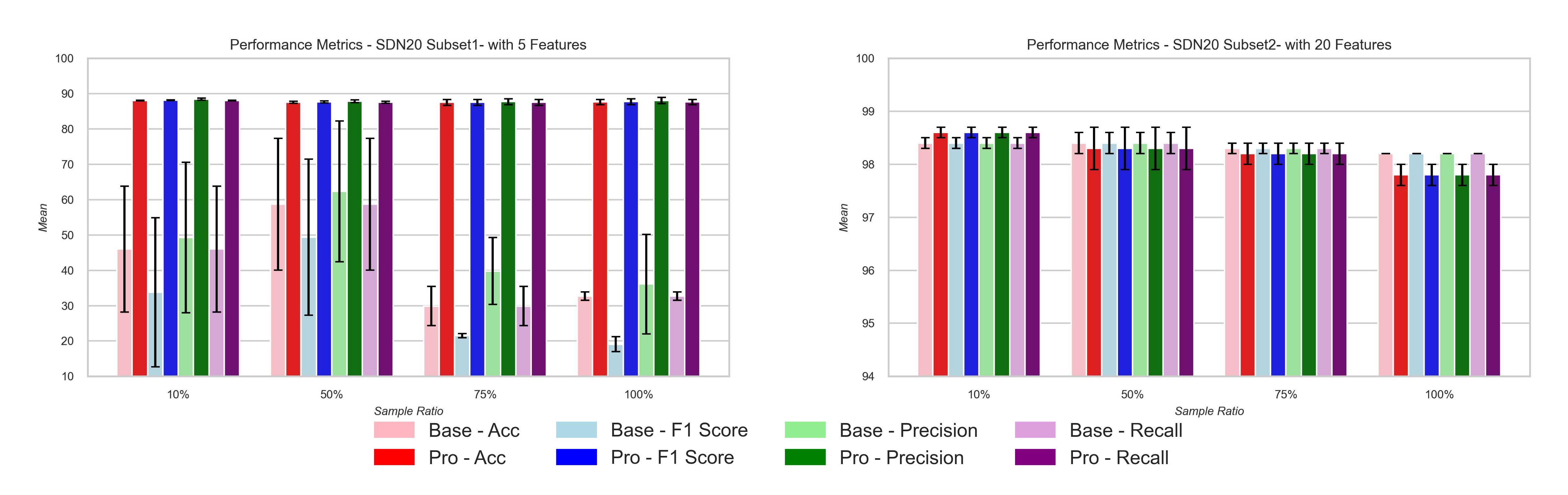}
    \caption{Evaluation of two test subsets from SDN20 dataset with 5 and 20 unique feature sets - Table \ref{tab:Com_table} (Sr\# 12)}
    \label{fig:sdn20_3}
\end{figure*}

\begin{figure*}[ht]
    \centering
    \includegraphics[width=1\linewidth,height=0.4\linewidth,keepaspectratio=false]{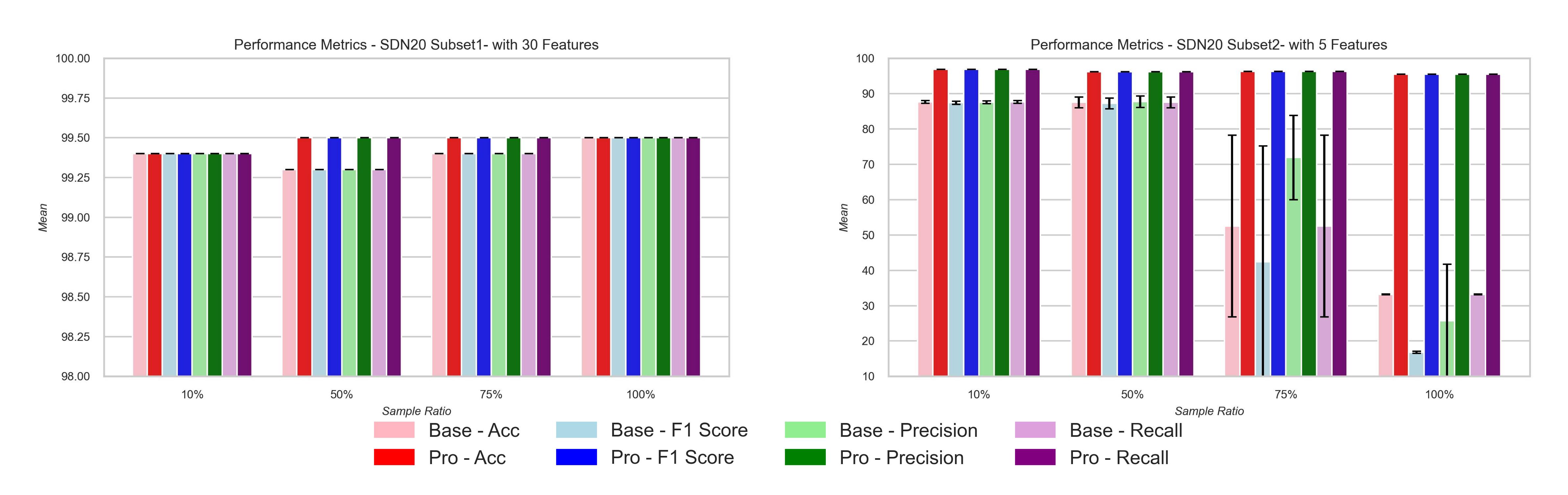}
    \caption{Evaluation of two test subsets from SDN20 dataset with 30 and 5 unique feature sets - Table \ref{tab:Com_table} (Sr\# 13)}
    \label{fig:sdn20_4}
\end{figure*}

In analyzing the figures and depicting subsets with an imbalance in the total number of features, as presented in Figure  \ref{fig:sdn20_3} (Subset-1 with 5 features and Subset-2 with 20 features) and Figure \ref{fig:sdn20_4} (Subset-1 with 30 features and Subset-2 with 5 features), the proposed model demonstrates significant improvement over the individual baseline classifiers. This improvement, as seen with the previous datasets, is its ability to effectively learn from both distinct subsets despite the differences in total number of feature sets. The proposed model is capable of aligning the distributions between these heterogeneous subsets, thereby resulting in improved classification performance for both test datasets.\\

In scenarios where one subset contains more features, such as Subset-2 in Figure \ref{fig:sdn20_3} or Subset-1 in Figure \ref{fig:sdn20_4}, the proposed model significantly benefits from leveraging this richer feature set, leading to improved scores across all evaluated metrics, including for the subset with fewer features, regardless of the variability in sample sizes. These scores illustrate the robustness of the proposed model in managing feature imbalance, effectively improving overall classification performance.\\

\subsection{Setup 2}
\subsubsection{Evaluation of combinatorial sets derived from two different datasets}

In this phase of the evaluation, we consider the combination of subsets derived from two entirely different datasets. Each subset is created by randomly selecting unique feature sets from the respective datasets. The performance of the proposed model is evaluated by applying it to these subsets and comparing the results against their individual baseline models. Here, we aim to assess the model's ability to generalize across heterogeneous data sources. This experiment allows us to investigate the model's robustness and effectiveness in aligning distributions and learning from diverse data subsets to better classify network samples.

\begin{figure}[]
    \centering\includegraphics[width=0.9\linewidth,height=0.6\linewidth,keepaspectratio=false]{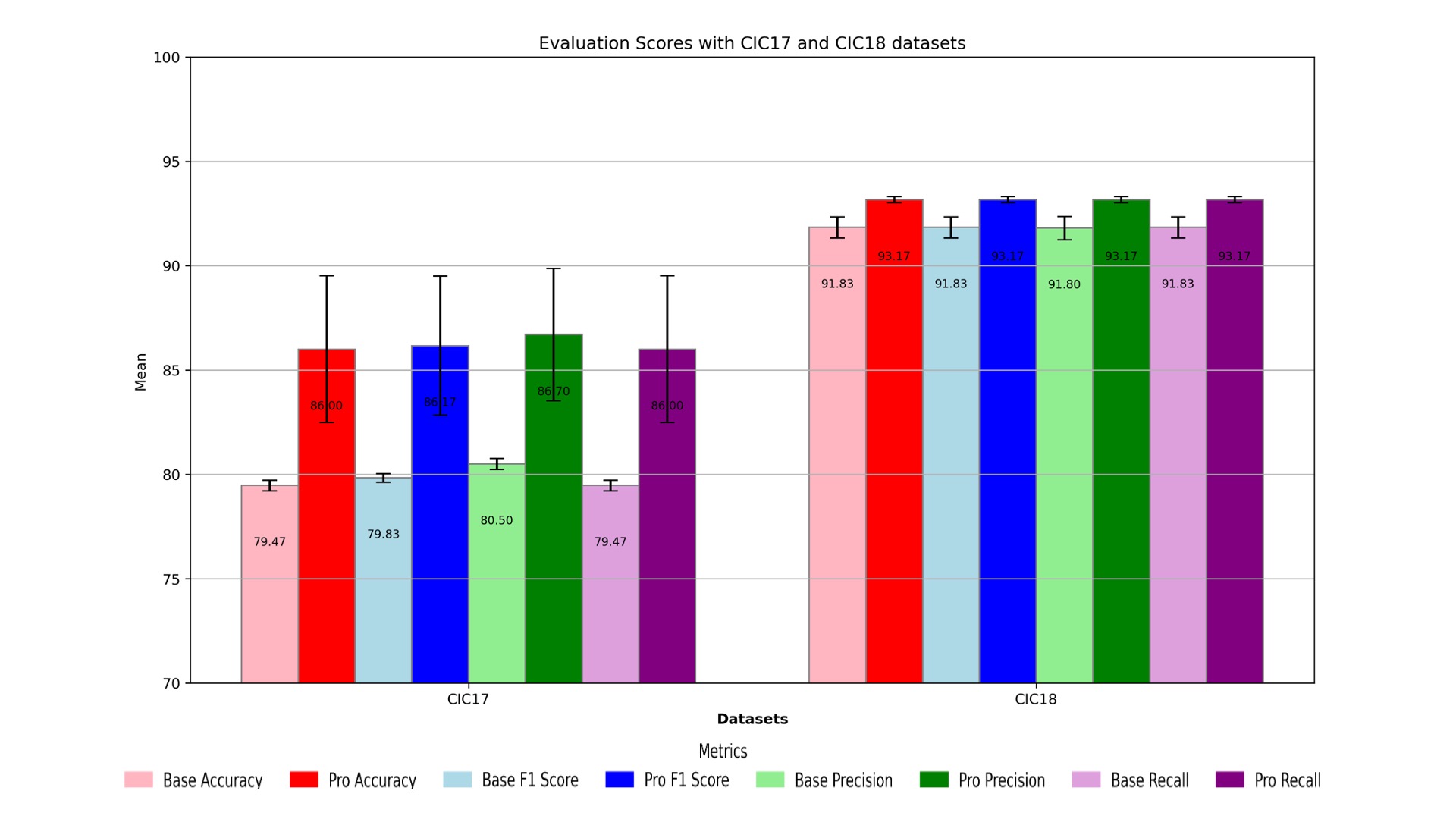}
    \caption{Evaluation of two test subsets from CIC-IDS2017 and CIC-IDS2018 dataset with 10  unique feature sets - Table \ref{tab:Com_table} (Sr\# 14). The evaluation plots compare the metrics from the proposed model and the respective baseline model.}
    \label{fig:cic17_cic18}
\end{figure}

Figure \ref{fig:cic17_cic18} presents the evaluation of the proposed model compared to the baseline models using subsets from the CIC-IDS2017 and CIC-IDS2018 datasets, each with 10 unique feature sets. The plots display the mean and standard deviation of the metrics, and it is evident that the proposed model consistently outperforms the baseline models across all metrics. The baseline accuracy is 79\% for CIC17 and 91\% for CIC18, while the proposed model achieves significantly higher accuracies of 86\% and 93\%, respectively. Similarly, F1 score, precision, and recall show notable improvements with the proposed model. Although the CIC17 plots exhibit higher variability compared to CIC18, the proposed model demonstrates a clear trend of improving classification performance. This highlights the robustness of the proposed model, even when dealing with heterogeneous datasets.\\

\begin{figure}[]
    \centering
    \includegraphics[width=0.9\linewidth,height=0.6\linewidth,keepaspectratio=false]{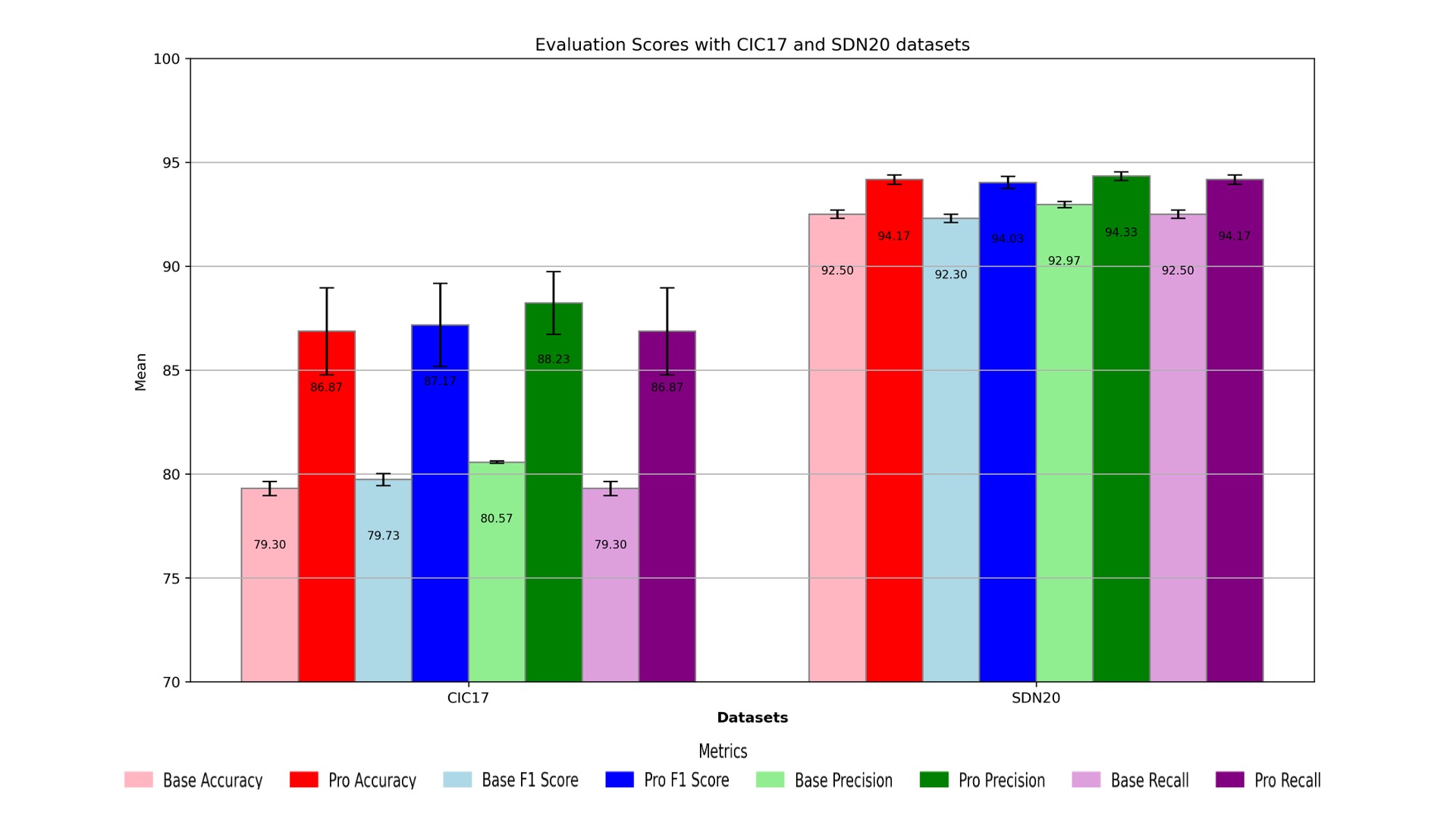}
    \caption{Evaluation of two test subsets from CIC-IDS2017 and SDN20 dataset with 10  unique feature sets - Table \ref{tab:Com_table} (Sr\# 15). The evaluation plots compare the metrics from the proposed model and the respective baseline model.}
    \label{fig:cic17_sdn20}
\end{figure}

The plots in Figure \ref{fig:cic17_sdn20} present the evaluation of the proposed model compared to baseline models using subsets from the CIC-IDS2017 and SDN20 datasets. Across both test datasets, the proposed model consistently outperforms the baseline models across all metrics. For CIC-IDS2017, the baseline accuracy is 79.30\%, while the proposed model achieves a significantly higher accuracy of 86.87\%. Similarly, the F1 score, precision, and recall show notable improvements, surpassing the baseline evaluation. For SDN20, the baseline accuracy is 92.50\%, and the proposed model increases this to 94.17\%, showing improvements across all other metrics as well. There is a notable variability in the scores with CIC-IDS2017, suggesting fluctuations in performance across different runs. However, the proposed model still demonstrates clear superiority over the baseline models. These results, derived from a combination of diverse test datasets, demonstrate the ability of the proposed model to efficiently handle heterogeneous data, align feature distributions, and improve the classification outcomes compared to the baseline models.\\

\begin{figure}[]
    \centering
    \includegraphics[width=0.9\linewidth,height=0.6\linewidth,keepaspectratio=false]{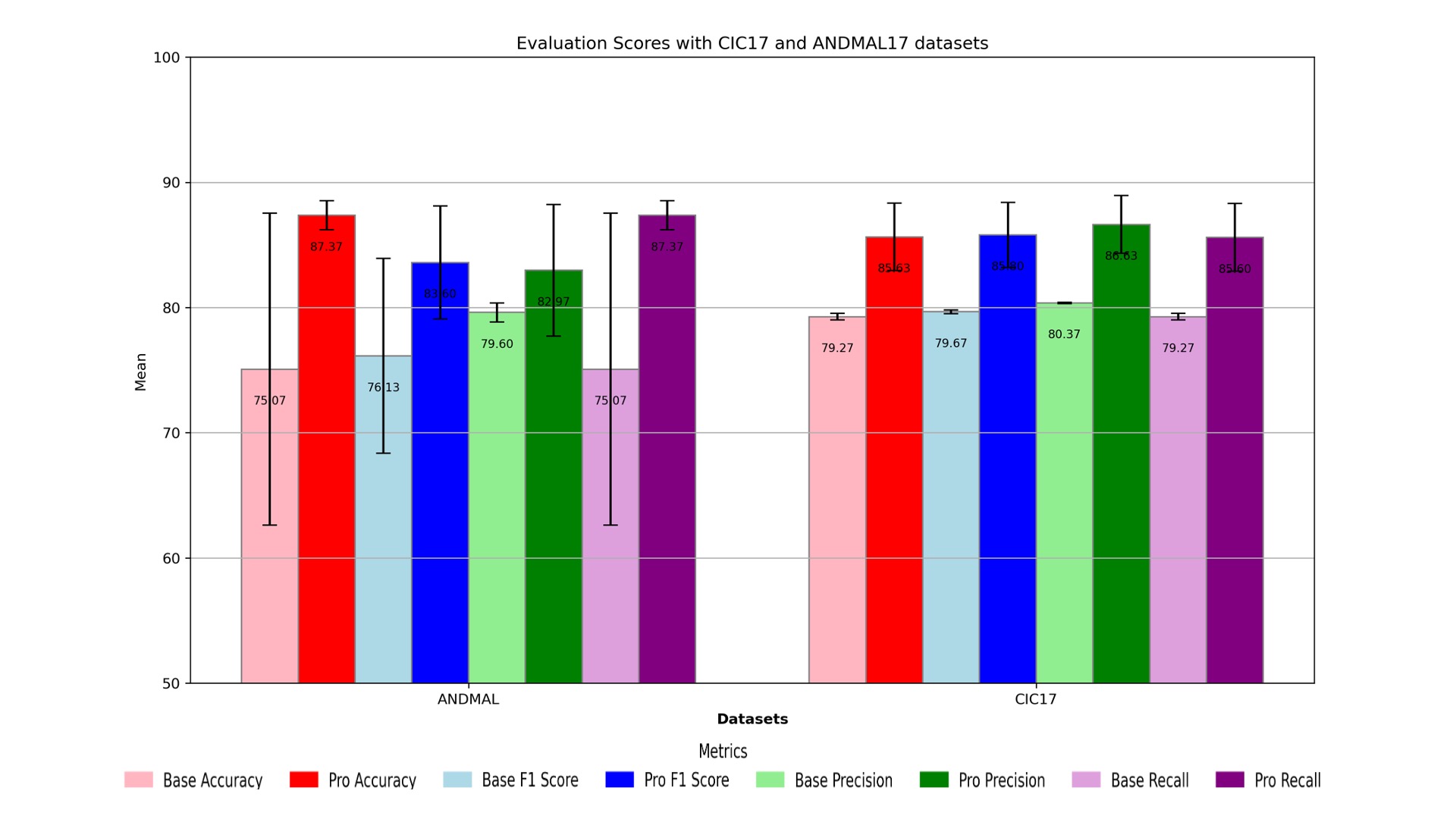}
    \caption{Evaluation of two test subsets from CIC-IDS2017 and ANDMAL2017 dataset with 10  unique feature sets - Table \ref{tab:Com_table} (Sr\# 16). The evaluation plots compare the metrics from the proposed model and the respective baseline model.}
    \label{fig:cic17_andmal}
\end{figure}

Figure \ref{fig:cic17_andmal} presents the evaluation of the proposed model using subsets from the ANDMAL2017 and CIC-IDS2017 datasets, showing the mean and standard deviation for each metric. For ANDMAL2017, the proposed model demonstrates significant improvement over the baseline, with accuracy increasing from 75.07\% to 87.37\%. Similar improvements are observed in the F1 score, precision, and recall. However, the higher variability in the ANDMAL2017 results, indicated by larger error bars, suggests variability in performance, possibly due to the quality of samples and label imbalances within the dataset, as discussed in Appendix \ref{appendixA}. For CIC-IDS2017, the baseline accuracy is 79.27\%, and the proposed model improves this to 85.63\%, with more consistent results and lower variability compared to ANDMAL2017. These improvements across both datasets highlight the robustness of the proposed model in aligning feature distributions and enhancing classification performance, even with diverse and imbalanced datasets.\\

\begin{figure}[]
    \centering
    \includegraphics[width=0.9\linewidth,height=0.6\linewidth,keepaspectratio=false]{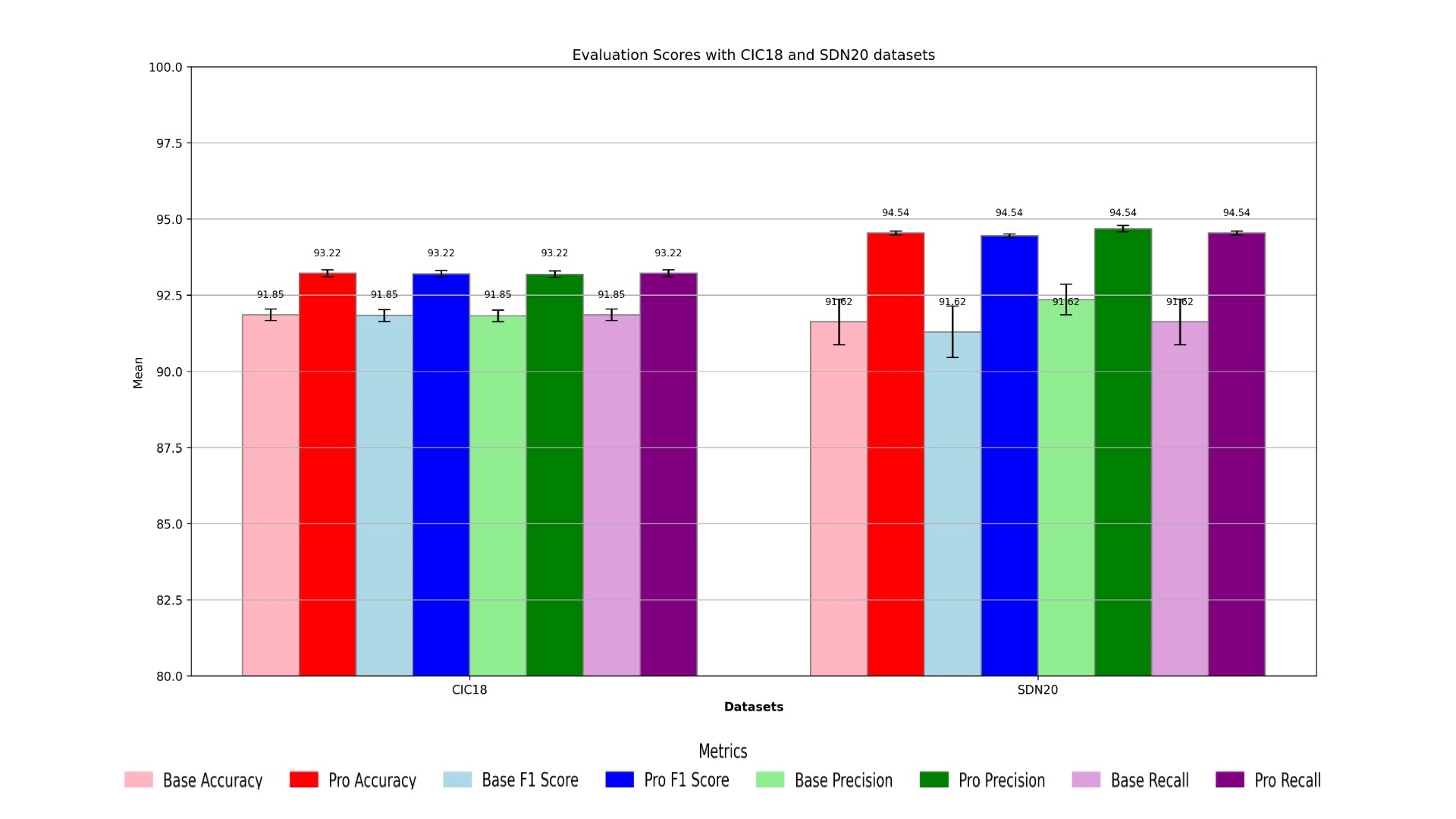}
    \caption{Evaluation of two test subsets from SDN20 and CIC-IDS2018 dataset with 10  unique feature sets - Table \ref{tab:Com_table} (Sr\# 17). The evaluation plots compare the metrics from the proposed model and the respective baseline model.}
    \label{fig:cic18_sdn20}
\end{figure}

Figure \ref{fig:cic18_sdn20} presents an evaluation of the proposed model using subsets from the CIC-IDS2018 and SDN20 datasets, each containing randomly selected 10 unique feature sets. For CIC-IDS2018, the baseline accuracy is 91.85\%, while the proposed model improves this to 93.22\%, showing a clear enhancement in performance across all metrics with relatively low variability. For SDN20, the baseline accuracy is 91.62\%, and the proposed model increases this to 94.54\%, with less variability in the evaluation, indicating more stable and consistent results. Overall, the proposed model demonstrates superior performance for both CIC-IDS2018 and SDN20, highlighting its robustness and ability to align feature distributions effectively, leading to better classification outcomes across heterogeneous datasets.\\

\begin{figure}[]
    \centering
    \includegraphics[width=0.9\linewidth,height=0.6\linewidth,keepaspectratio=false]{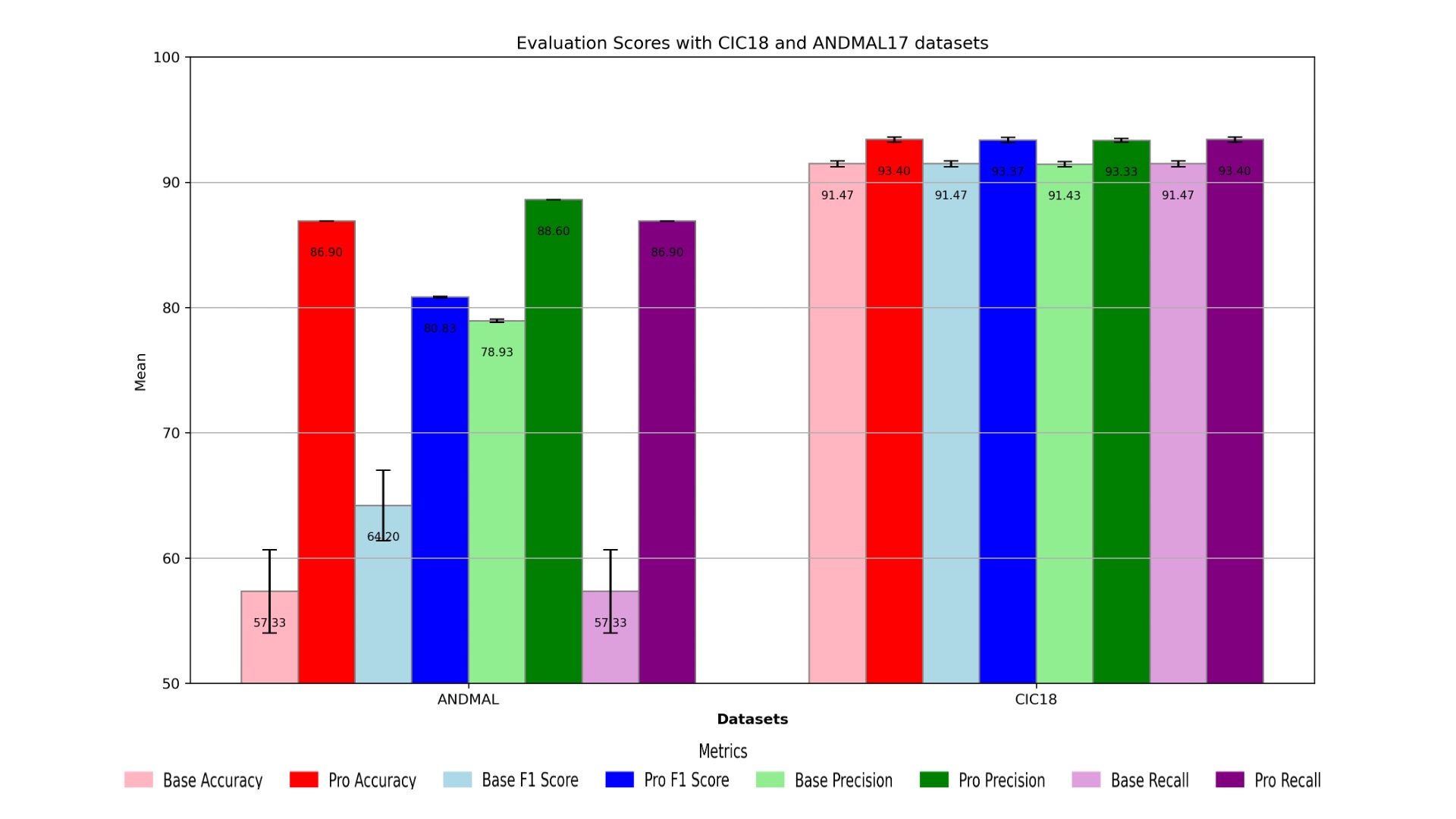}
    \caption{Evaluation of two test subsets from ANDMAL and CIC-IDS2018 dataset with 10  unique feature sets - Table \ref{tab:Com_table} (Sr\# 18). The evaluation plots compare the metrics from the proposed model and the respective baseline model.}
    \label{fig:cic18_andmal}
\end{figure}

figure \ref{fig:cic18_andmal} presents an evaluation of the proposed model using randomly selected subsets from the ANDMAL2017 and CIC-IDS2018 datasets. For ANDMAL2017, the baseline accuracy is as low as 57.33\%, while the proposed model significantly improves this to 86.90\%, demonstrating an enhancement in performance across all metrics. However, the higher variability in the results may be attributed to the quality of the data and label imbalances to have a biased result, which was similarly observed in the evaluation with CIC-IDS2017-ANDMAL (Table \ref{tab:Com_table} - Sr\# 16). For CIC-IDS2018, the baseline accuracy is 91.47\%, and the proposed model increases this to 93.40\%, showing more stable and consistent performance when compared with ANDMAL2017 scores. These results demonstrate that the proposed model tries to achieve a synergistic improvement, enhancing the detection of samples from both diverse datasets and improving overall classification performance. The findings emphasize the model's robustness and ability to align feature distributions, delivering better classification outcomes across heterogeneous datasets.\\

\begin{figure}[]
    \centering
    \includegraphics[width=0.9\linewidth,height=0.5\linewidth,keepaspectratio=false]{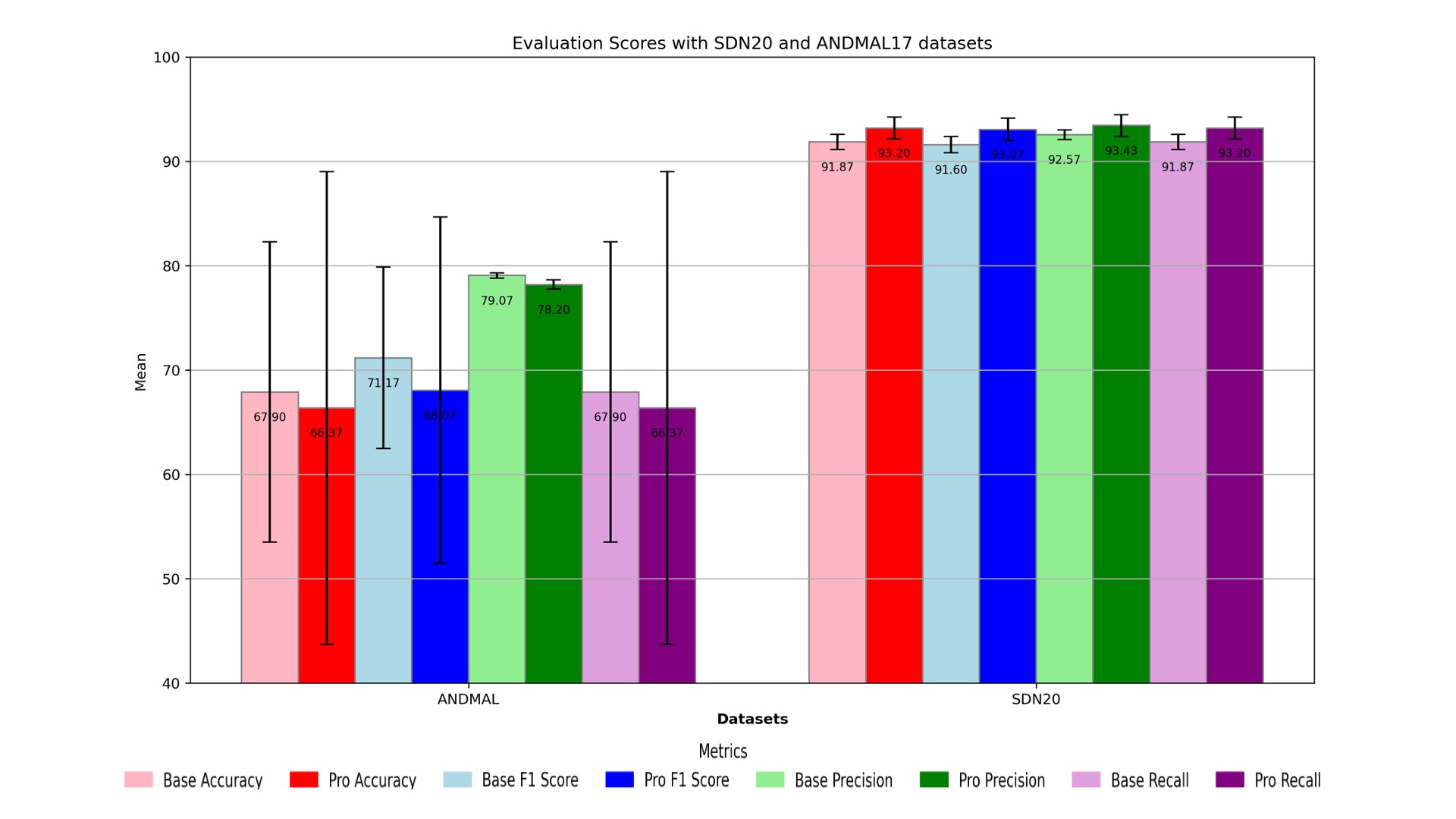}
    \caption{Evaluation of two test subsets from ANDMAL and SDN20 dataset with 10  unique feature sets - Table \ref{tab:Com_table} (Sr\# 19). The evaluation plots compare the metrics from the proposed model and the respective baseline model.}
    \label{fig:andmal_sdn20}
\end{figure}

Our final combination of two diverse datasets, ANDMAL2017 and SDN20, as illustrated in Figure \ref{fig:andmal_sdn20}, presents the evaluation of the proposed model in comparison to the baseline scores. Interestingly, for ANDMAL2017, the baseline accuracy exceeds that of the proposed model, which underperforms with an accuracy of 66.37\%, a trend that is similar across F1 score, precision, and recall. Additionally, the results for ANDMAL2017 exhibit significant variability, indicating inconsistent performance across repeated runs. This variability could be attributed to data quality issues and label imbalances in the ANDMAL dataset, which were also observed in previous analyses, both when evaluated individually and when combined with other datasets. In contrast, for the SDN20 dataset, the proposed model outperforms the baseline model across all metrics, demonstrating minimal variability and consistent performance across repeated experiments. These results suggest that, while the proposed model generally performs well and shows robustness with SDN20, it struggles with datasets like ANDMAL2017, which contain complex or imbalanced data. Despite these challenges, the proposed model still demonstrates its effectiveness in improving classification performance, particularly with more balanced and stable datasets.\\

\subsection{Setup 3}
\subsubsection{Evaluation of combinatorial sets derived from three different datasets}
\begin{figure}[]
    \centering
    \includegraphics[width=1\linewidth]{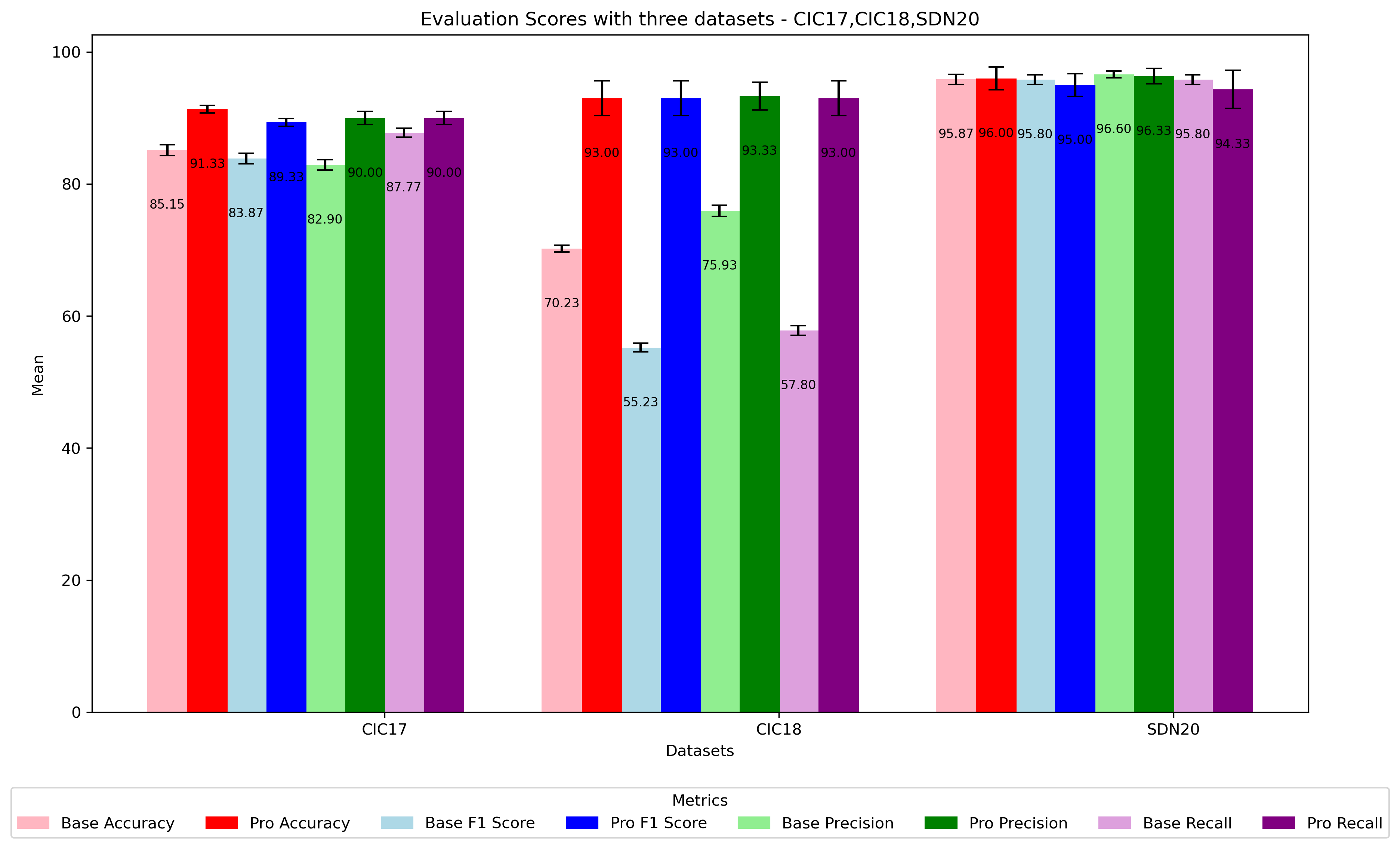}
    \caption{Comparison of evaluation scores between the baseline and proposed models across three distinct datasets with unique feature sets  from` CICIDS17, CICIDS18 and SDN20 -  Table \ref{tab:Com_table} (Sr\# 20).}
    \label{fig:CIC17_CIC18_SDN20}
\end{figure}

To further evaluate the capability of the proposed model in handling multiple datasets, we introduced three different datasets for testing. Given that the model is designed to adapt to datasets from different domains, each representing varied sample distributions, the objective here is to assess the model's effectiveness in learning from unique feature sets across three different datasets and its ability to generalize across diverse distributions. Figure \ref{fig:CIC17_CIC18_SDN20} presents the performance evaluation scores of the proposed model on the test subsets compared to their respective baseline scores.\\

The figure presents an evaluation of the proposed model using subsets from three datasets: CIC-IDS2017, CIC-IDS2018, and SDN20. These datasets were selected due to their balanced nature and higher data quality compared to ANDMAL, as observed in previous experiments. Three subsets with 10 unique feature sets were extracted, as detailed in Table \ref{tab:Com_table} (Sr\# 20). The proposed model improves accuracy from 85.15\% to 91.33\% for CIC-IDS2017, with similar improvements in F1 score, precision, and recall. For CIC-IDS2018, the baseline accuracy is 70.23\%, while the proposed model achieves a significantly higher accuracy of 93\%, along with substantial improvements in F1 score, precision, and recall. In the case of SDN20, both the baseline and proposed models exhibit high accuracy, with the baseline at 95.87\% and the proposed model slightly improving this to 96\%. All other metrics for SDN20 show minimal variability, indicating stable performance.

Overall, these results demonstrate that the proposed model consistently outperforms the baseline across all three datasets, where it shows significant classification improvements. The evaluation validates that the proposed model adapts well to heterogeneous datasets, leading to enhanced classification performance and better detection of network samples.

\subsection{Monitoring the alignment of the domains in the latent space by MMD}
\begin{figure}[]
    \centering
    \includegraphics[width=1\linewidth]{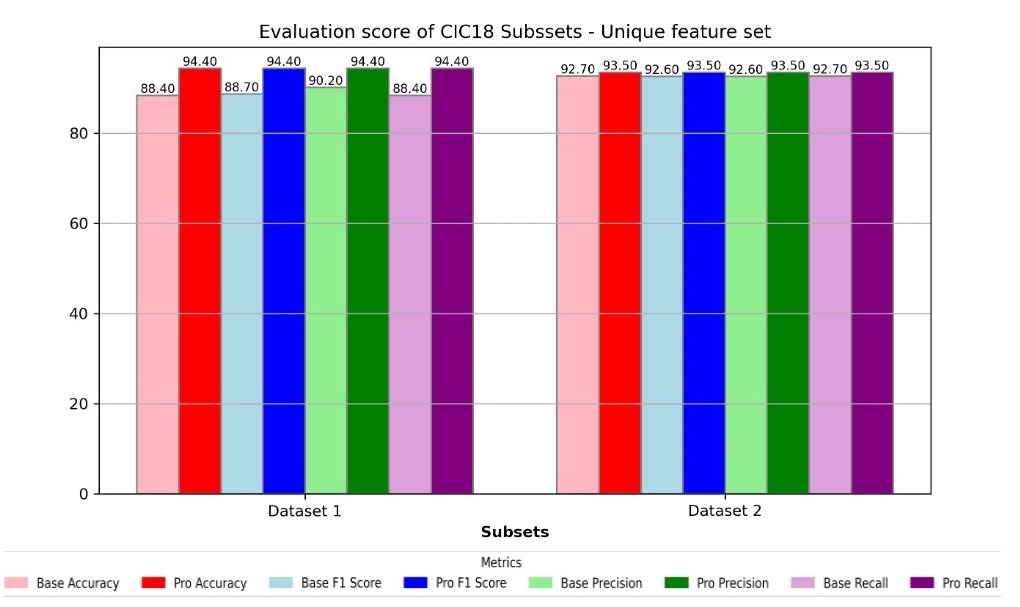}
    \caption{Comparison of evaluation scores between the baseline and proposed models across distinct subsets with 10 and 5 unique feature sets(Dataset1 - Dataset2) derived from the CIC-IDS18dataset - Table \ref{tab:Com_table} (Sr\# 1).}
    \label{fig:init_subsets_comp}
\end{figure}

\begin{figure}[ht]
    \centering
    \includegraphics[width=0.75\linewidth]{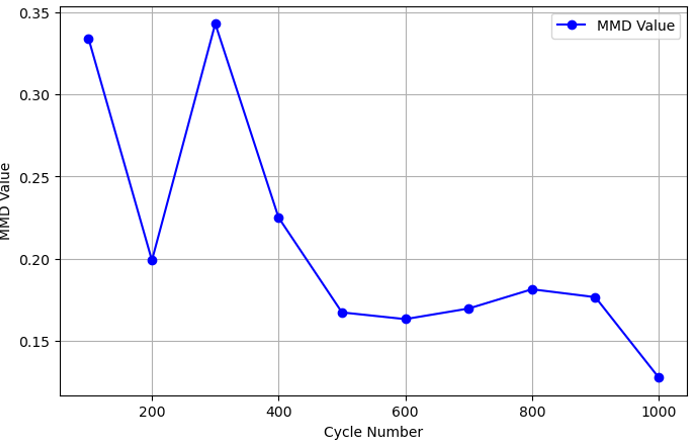}
    \caption{A plot illustrating the mean maximum discrepancy value over the entire training cycle for the CIC18 subsets, showing a decreasing trend that indicates the alignment of feature distributions.}
    \label{fig:MMD PLot}
\end{figure}

We extracted two subsets of features from the CIC-IDS2018 dataset, randomly selecting 10 and 5 unique feature sets to simulate a cross-domain scenario, as detailed in Table \ref{tab:Com_table} (Sr\# 1). The pre-processed subsets were first passed through the baseline classifier to evaluate classification scores and then through the proposed model. Figure \ref{fig:init_subsets_comp} illustrates the results of this experiment. From the plot, we observe that the proposed model effectively learns and adapts from both domains, improving network intrusion detection across them, as seen in the increasing trend of the classification scores in figure \ref{fig:init_subsets_comp}. We hypothesize that this improvement is due to the minimization of the Maximum Mean Discrepancy (MMD), leading to better alignment between the datasets. Figure \ref{fig:MMD PLot} supports this hypothesis, showing a decreasing trend in MMD throughout the training process of the proposed model.

\section{Conclusion} 
\label{ch:Results} 
This paper presents a novel neural network architecture for network intrusion detection, addressing the challenge of handling heterogeneous datasets. The proposed model is designed to integrate multiple domain-specific feature sets, enabling it to learn domain-invariant representations from varying input features, samples, and distributions. This design aims to enhance the accuracy and generalizability of intrusion detection and classification systems.

The core of the model is a feature extraction mechanism that incorporates multimodal learning within a domain adaptation framework. This allows the model to effectively capture domain-specific variations while aligning distributions across domains. To evaluate the model, we conducted four experiments: (1) testing on two distinct feature subsets from a single dataset, (2) analyzing the effect of shared features across subsets, (3) using random feature subsets, and (4) evaluating performance on multiple datasets with varying sample sizes and distributions.

\ifCLASSOPTIONcompsoc
  \section*{Acknowledgments}
\else
  \section*{Acknowledgment}
\fi
This research has been funded by the Vinnova, Strategic Vehicle
Research and Innovation programme.

\bibliographystyle{elsarticle-num} 
\bibliography{References}

\appendices
\label{appendixA}
\section{}

\begin{table*}[]

\centering
\caption{Comprehensive list of 78 network traffic features that are consistently present across four intrusion detection datasets: CIC-IDS2017, CIC-IDS2018, SDN 2020, and ANDMAL.}
\label{Dataset_Feature}
\begin{tabular}{@{}clcll@{}}
\toprule
\textbf{Feat ID} & \multicolumn{1}{c}{\textbf{Feature Name}} & \textbf{Feat ID} & \multicolumn{1}{c}{\textbf{Feature Name}} &  \\ \midrule
1  & Protocol                        & 40 & Max\_Packet\_Length        &  \\
2  & Flow\_Duration                  & 41 & Packet\_Length\_Mean       &  \\
3  & Total\_Fwd\_Packets             & 42 & Packet\_Length\_Std        &  \\
4  & Total\_Backward\_Packets        & 43 & Packet\_Length\_Variance   &  \\
5  & Total\_Length\_of\_Fwd\_Packets & 44 & FIN\_Flag\_Count           &  \\
6  & Total\_Length\_of\_Bwd\_Packets & 45 & SYN\_Flag\_Count           &  \\
7  & Fwd\_Packet\_Length\_Max        & 46 & RST\_Flag\_Count           &  \\
8  & Fwd\_Packet\_Length\_Min        & 47 & PSH\_Flag\_Count           &  \\
9  & Fwd\_Packet\_Length\_Mean       & 48 & ACK\_Flag\_Count           &  \\
10 & Fwd\_Packet\_Length\_Std        & 49 & URG\_Flag\_Count           &  \\
11 & Bwd\_Packet\_Length\_Max        & 50 & CWE\_Flag\_Count           &  \\
12 & Bwd\_Packet\_Length\_Min        & 51 & ECE\_Flag\_Count           &  \\
13 & Bwd\_Packet\_Length\_Mean       & 52 & DownUp\_Ratio              &  \\
14 & Bwd\_Packet\_Length\_Std        & 53 & Average\_Packet\_Size      &  \\
15 & Flow\_Bytess                    & 54 & Avg\_Fwd\_Segment\_Size    &  \\
16 & Flow\_Packetss                  & 55 & Avg\_Bwd\_Segment\_Size    &  \\
17 & Flow\_IAT\_Mean                 & 56 & Fwd\_Avg\_BytesBulk        &  \\
18 & Flow\_IAT\_Std                  & 57 & Fwd\_Avg\_PacketsBulk      &  \\
19 & Flow\_IAT\_Max                  & 58 & Fwd\_Avg\_Bulk\_Rate       &  \\
20 & Flow\_IAT\_Min                  & 59 & Bwd\_Avg\_BytesBulk        &  \\
21 & Fwd\_IAT\_Total                 & 60 & Bwd\_Avg\_PacketsBulk      &  \\
22 & Fwd\_IAT\_Mean                  & 61 & Bwd\_Avg\_Bulk\_Rate       &  \\
23 & Fwd\_IAT\_Std                   & 62 & Subflow\_Fwd\_Packets      &  \\
24 & Fwd\_IAT\_Max                   & 63 & Subflow\_Fwd\_Bytes        &  \\
25 & Fwd\_IAT\_Min                   & 64 & Subflow\_Bwd\_Packets      &  \\
26 & Bwd\_IAT\_Total                 & 65 & Subflow\_Bwd\_Bytes        &  \\
27 & Bwd\_IAT\_Mean                  & 66 & Init\_Win\_bytes\_forward  &  \\
28 & Bwd\_IAT\_Std                   & 67 & Init\_Win\_bytes\_backward &  \\
29 & Bwd\_IAT\_Max                   & 68 & act\_data\_pkt\_fwd        &  \\
30 & Bwd\_IAT\_Min                   & 69 & min\_seg\_size\_forward    &  \\
31 & Fwd\_PSH\_Flags                 & 70 & Active\_Mean               &  \\
32 & Bwd\_PSH\_Flags                 & 71 & Active\_Std                &  \\
33 & Fwd\_URG\_Flags                 & 72 & Active\_Max                &  \\
34 & Bwd\_URG\_Flags                 & 73 & Active\_Min                &  \\
35 & Fwd\_Header\_Length\_2          & 74 & Idle\_Mean                 &  \\
36 & Bwd\_Header\_Length             & 75 & Idle\_Std                  &  \\
37 & Fwd\_Packetss                   & 76 & Idle\_Max                  &  \\
38 & Bwd\_Packetss                   & 77 & Idle\_Min                  &  \\
39 & Min\_Packet\_Length             & 78 & multilabel                 &  \\ \bottomrule
\end{tabular}
\end{table*}

\begin{table*}[]
\centering
\caption{Feature ID combinations selected from four different network intrusion datasets (CICIDS2017, CICIDS2018, SDN-2020, ANDMAL) for validating the proposed novel model. Sr\#1 to Sr\#13 represent combinations where features were randomly selected from a single individual dataset. Sr\#14 to Sr\#19 involve combinations drawn from two different datasets. The table lists the exact Feature IDs used in each combination for reproducibility and further analysis.}
\resizebox{1\textwidth}{0.4\textheight}{%
    \begin{tabular}{c|l|c|c|c|l}
    \toprule
    \textbf{Sr\#} &
      \multicolumn{1}{c|}{\textbf{Dataset/s}} &
      \textbf{Feature   Combination} &
      \textbf{Dataset 1 (Feat. Id)} &
      \textbf{Dataset 2 (Feat. Id)} &
      \multicolumn{1}{c|}{\textbf{Dataset 3 (Feat. Id)}} \\ \midrule
    1 &
      CIC18 &
      10-5 &
      \begin{tabular}[c]{@{}c@{}}1,10,11,24,34,35\\ ,40,50,73,77\end{tabular} &
      9,37,52,62,65 &
       \\ \midrule
    2 &
      \multirow{4}{*}{CIC18} &
      5-5 &
      13,20,26,33,72 &
      3,7,25,37,64 &
       \\ 
       \cline{1-1} \cline{3-6} 
    3 &
       &
      15-15 &
      \begin{tabular}[c]{@{}c@{}}4,5,7,10,15,19,20,21,\\ 22,24,26,41,42,47,57\end{tabular} &
      \begin{tabular}[c]{@{}c@{}}13,14,17,23,27,35,\\ 37,48,50,56,59,\\ 68,71,73,77\end{tabular} &
       \\ \cline{1-1} \cline{3-6} 
    4 &
       &
      5-20 &
      14,24,45,68,73 &
      \begin{tabular}[c]{@{}c@{}}1,10,13,15,16,21,\\ 22,23,29,31,38,41,42,\\ 56,58,60,64,65,70,72\end{tabular} &
       \\ \cline{1-1} \cline{3-6} 
    5 &
       &
      30-5 &
      \begin{tabular}[c]{@{}c@{}}1,2,3,5,8,13,17,19,20,\\ 21,25,31,39,40,42,46,47,\\ 53,56,57,58,61,62,65,\\ 66,67,70,74,76,77\end{tabular} &
      24,29,30,41,44,78 &
       \\ \midrule 
    6 &
      \multirow{4}{*}{CIC17} &
      5-5 &
      12,14,18,53.70 &
      10,22,30,54,73 &
       \\ \cline{1-1} \cline{3-6} 
    7 &
       &
      15-15 &
      \begin{tabular}[c]{@{}c@{}}9,11,15,22,32,35,\\ 41,45,47,53,54,\\ 56,58,65,67\end{tabular} &
      \begin{tabular}[c]{@{}c@{}}13,14,16,28,34,40,\\ 44,51,59,60,64,\\ 68,71,72,74\end{tabular} &
       \\ \cline{1-1} \cline{3-6} 
    8 &
       &
      5-20 &
      29,62,71,75,77 &
      \begin{tabular}[c]{@{}c@{}}6,15,24,25,27,35,41,\\ 44,47,49,50,52,53,\\ 61,64,66,67,\\ 69,74,76\end{tabular} &
       \\ \cline{1-1} \cline{3-6} 
    9 &
       &
      30-5 &
      \begin{tabular}[c]{@{}c@{}}1,2,3,4,5,6,7,15,16,18,\\ 20,22,23,24,25,29,32,\\ 36,37,39,43,45,49,\\ 51,57,62,65,69,74,75\end{tabular} &
      8,31,33,61,67 &
       \\ \midrule
    10 &
      \multirow{4}{*}{SDN20} &
      5-5 &
      6,27,30,36,69 &
      13,15,20,62,66 &
       \\ \cline{1-1} \cline{3-6} 
    11 &
       &
      15-15 &
      \begin{tabular}[c]{@{}c@{}}5,6,16,19,20,21,\\ 23,24,26,33,52,\\ 53,65,75,77\end{tabular} &
      \begin{tabular}[c]{@{}c@{}}2,3,9,13,14,29,\\ 34,41,51,60,67,\\ 71,73,74,76\end{tabular} &
       \\ \cline{1-1} \cline{3-6} 
    12 &
       &
      5-20 &
      2,31,35,70,73 &
      \begin{tabular}[c]{@{}c@{}}1,11,12,14,15,16,17,\\ 18,20,24,28,33,44,\\ 50,54,56,66,\\ 69,71,77\end{tabular} &
       \\ \cline{1-1} \cline{3-6} 
    13 &
       &
      30-5 &
      \begin{tabular}[c]{@{}c@{}}2,4,6,8,14,18,19,23,\\ 25,26,30,36,37,38,40,\\ 44,45,52,55,56,57,59,\\ 61,63,65,70,73,74,75,77\end{tabular} &
      11,15,29,33,47 &
       \\ \midrule
    14 &
      CIC17-CIC18 &
      10-10 &
      \begin{tabular}[c]{@{}c@{}}5,12,15,25,33,51,\\ 64,66,67,77\end{tabular} &
      \begin{tabular}[c]{@{}c@{}}11,17,26,29,31,\\ 32,37,45,56,73\end{tabular} &
       \\ \midrule
    15 &
      CIC17-SDN20 &
      10-10 &
      \begin{tabular}[c]{@{}c@{}}5,12,15,25,33,\\ 51,64,66,67,77\end{tabular} &
      \begin{tabular}[c]{@{}c@{}}1,6,7,9,28,54,\\ 55,63,65,74\end{tabular} &
       \\ \midrule
    16 &
      CIC17-ANDMAL &
      10-10 &
      \begin{tabular}[c]{@{}c@{}}5,12,15,25,33,\\ 51,64,66,67,77\end{tabular} &
      \begin{tabular}[c]{@{}c@{}}4,22,35,42,43,46,\\ 48,50,69,71\end{tabular} &
       \\ \midrule
    17 &
      CIC18-SDN20 &
      10-10 &
      \begin{tabular}[c]{@{}c@{}}11,17,26,29,31,\\ 32,37,45,56,73\end{tabular} &
      \begin{tabular}[c]{@{}c@{}}1,6,7,9,28,54,\\ 55,63,65,74\end{tabular} &
       \\ \midrule
    18 &
      CIC18-ANDMAL &
      10-10 &
      \begin{tabular}[c]{@{}c@{}}11,17,26,29,31,\\ 32,37,45,56,73\end{tabular} &
      \begin{tabular}[c]{@{}c@{}}4,22,35,42,43,\\ 46,48,50,69,71\end{tabular} &
       \\ \midrule
    19 &
      SDN20-ANDMAL &
      10-10 &
      \begin{tabular}[c]{@{}c@{}}1,6,7,9,28,54,\\ 55,63,65,74\end{tabular} &
      \begin{tabular}[c]{@{}c@{}}4,22,35,42,43,\\ 46,48,50,69,71\end{tabular} &
       \\ \midrule
    20 &
      CIC17-CIC18-SDN20 &
      10-10-10 &
      \begin{tabular}[c]{@{}c@{}}1,3,7,10,18,26,\\ 38,48,54,55,56,\\ 68,69,71,75\end{tabular} &
      \begin{tabular}[c]{@{}c@{}}6,7,9,10,15,16,\\ 19,21,22,27,37,\\ 40,46,48,58,\\ 70,74\end{tabular} &
      \multicolumn{1}{c}{\begin{tabular}[c]{@{}c@{}}8,10,15,17,21,\\ 34,38,51,55,56,\\ 59,63,67,69,73\end{tabular}} \\ \bottomrule
    \end{tabular}%
    }
    
    \label{tab:Com_table}
    \end{table*}

\ifCLASSOPTIONcompsoc
\else
\fi

\ifCLASSOPTIONcaptionsoff
  \newpage
\fi

\end{document}